\documentclass[aps,prd,floats,floatfix,twocolumn,nofootinbib,amssymb]{revtex4}

\usepackage{graphicx}
\usepackage{bm}

\newcommand{\half}{{\scriptstyle\frac{1}{2}}}
\newcommand{\fourth}{{\scriptstyle\frac{1}{4}}}

\newcommand{\beq}{\begin{equation}}
\newcommand{\eeq}{\end{equation}}
\newcommand{\beqa}{\begin{eqnarray}}
\newcommand{\eeqa}{\end{eqnarray}}

\usepackage[usenames]{color}
\usepackage{xcolor}
\usepackage{ulem}
\begin{document}

\title{An Improved Gauge Driver for the Generalized Harmonic Einstein System}

\author{Lee Lindblom and B\'ela Szil\'agyi}

\affiliation{Theoretical Astrophysics 350-17, California Institute of
Technology, Pasadena, CA 91125}

\begin{abstract}
A new gauge driver is introduced for the generalized harmonic (GH)
representation of Einstein's equation.  This new driver allows a
rather general class of gauge conditions to be implemented in a way
that maintains the hyperbolicity of the combined evolution system.
This driver is more stable and effective, and unlike previous drivers,
allows stable evolutions using the dual-frame evolution technique.
Appropriate boundary conditions for this new gauge driver are
constructed, and a new boundary condition for the ``gauge'' components
of the spacetime metric in the GH Einstein system is introduced.  The
stability and effectiveness of this new gauge driver are demonstrated
through numerical tests, which impose a new damped-wave gauge condition
on the evolutions of single black-hole spacetimes.
\end{abstract}


\date{\today}

\maketitle

%%%%%%%%%%%%%%%%%%%%%%%%%%%%%%%%%%%%%%%%%%%%%%%%%%%%%%%%%%%%%%%%%%%%%%%%%%%%%%%
\section{Introduction}
\label{s:Introduction}
%%%%%%%%%%%%%%%%%%%%%%%%%%%%%%%%%%%%%%%%%%%%%%%%%%%%%%%%%%%%%%%%%%%%%%%%%%%%%%%

The gauge (or coordinate) degrees of freedom in the generalized
harmonic (GH) form of the Einstein equations are determined by specifying
the gauge-source functions $H^a$.  These functions are defined as the
results of the co-variant scalar-wave operator acting on each of the
spacetime coordinates $x^a$:
\begin{equation}
 H^a=\nabla^c\nabla_c\, x^a.
\label{e:GaugeSourceDef}
\end{equation}
(We use Latin letters from the beginning of the alphabet, $a$, $b$,
$c$, ...,  for spacetime indices.)  The GH form of Einstein's
equations can be represented (somewhat abstractly) as
\begin{equation}
\psi^{cd}\partial_c\partial_d \psi_{ab} + \partial_a H_b + \partial_b H_a=
Q_{ab}(H,\psi,\partial\psi),
\label{e:Einstein}
\end{equation}
where $\psi_{ab}$ is the spacetime metric, $H_a = \psi_{ab}H^b$, and
$Q_{ab}$ represents lower-order terms that depend on $H_a$, the
metric, and its first derivatives.  These equations are manifestly
hyperbolic whenever $H_a$ is specified as an explicit function of the
coordinates and the metric: $H_a = H_a(x,\psi)$.  In this case the
terms $\partial_a H_b$ appearing in Eq.~(\ref{e:Einstein}) contain
at most first derivatives of the metric.  The Einstein equations
become, therefore, a set of second-order wave equations for each
component of the spacetime metric:
\begin{equation}
\psi^{cd}\partial_c\partial_d\psi_{ab}=\hat Q_{ab}(x,\psi,\partial\psi).
\end{equation}
Thus the Einstein equations are manifestly hyperbolic for any $H_a =
H_a(x,\psi)$.

Most of the useful gauge conditions developed by the numerical
relativity community over the past several decades can not,
unfortunately, be expressed in the simple form $H_a =H_a(x,\psi)$,
unless the full spacetime metric $\psi_{ab}=\psi_{ab}(x)$ is known
{\it a priori}\hspace{0.1em}.  Many of these conditions (e.g.,
Bona-Mass\'o slicing or the $\Gamma$-driver shift conditions) would
require gauge-source functions that depend on the spacetime metric and
its first derivatives: $H_a=H_a(x,\psi,\partial\psi)$,
cf. Ref.\cite{Lindblom2007}.  In this case the terms $\partial_aH_b$
in Eq.~(\ref{e:Einstein}) would depend on the second derivatives of
the metric, $\psi_{ab}$, and this (generically) destroys the
hyperbolicity of the system.

This problem can be overcome by elevating $H_a$ to the status of an
independent dynamical field and introducing suitable evolution
equations for $H_a$, which we call gauge drivers~\cite{Lindblom2007,
Pretorius2005a, Pretorius2006}.  One obvious choice is to construct
gauge-driver equations that force $H_a$ to evolve toward the desired
gauge, e.g., $H_a\rightarrow F_a$ where $F_a$ is the target for the
selected gauge.  To be useful these gauge driver equations must also
make the combined Einstein gauge-driver system hyperbolic.  It is
fairly easy to construct hyperbolic evolution systems designed to
evolve $H_a$ toward any target $F_a(x,\psi,\partial\psi)$ that depends
on the spacetime metric and its first
derivatives~\cite{Lindblom2007}. Many of the gauge conditions found
most useful by the numerical relativity community have targets $F_a$
that belong to this class.  In most cases however, the coupled
Einstein gauge-driver evolution equations are unstable and the evolved
$H_a$ does not evolve robustly toward every target $F_a$ in this class
for generic evolutions.  The Einstein gauge-driver system is very
complicated, and there are many opportunities for unstable couplings
to develop between the dynamics of the spacetime metric and the
dynamics of the gauge field $H_a$.  Some gauge conditions, including
certain Bona-Mass\'o slicing conditions and some versions of the
$\Gamma$-driver shift conditions, have been implemented fairly
successfully using gauge drivers of this type in full 3D evolutions of
strongly perturbed single black-hole spacetimes~\cite{Lindblom2007}.
However, we find that even these ``successful'' gauge drivers fail
when more complicated simulations are attempted, e.g., evolving a
single black hole in a rotating reference frame or evolving black-hole
binary systems.

The purpose of this paper is to develop a better gauge driver that
overcomes some of these problems.  To this end we introduce in
Sec.~\ref{s:FirstOrderGaugeDriver} a new class of ``first-order''
gauge driver evolution equations, which are considerably simpler than
earlier drivers.  The dynamical simplicity of these new drivers
reduces the internal dynamical degrees of freedom available to $H_a$
(in a sense discussed in more detail in
Sec.~\ref{s:FirstOrderGaugeDriver}), hence reducing the possibility of
unwanted feedback or resonances with the dynamics of the Einstein
system.  We describe numerical tests of this new gauge-driver system
in Sec.~\ref{s:NumericalTests} that use a new damped-wave gauge
introduced in Appendix~\ref{s:DampedHarmonicGaugeConditions} to
provide an interesting non-trivial dynamical target $F_a$.  Using this
target $F_a$ we perform a series of numerical tests that evolve single
black-hole spacetimes with large dynamical gauge perturbations.  These
tests demonstrate the effectiveness and stability of the new
gauge-driver system for single- and dual-coordinate frame evolutions.
The strongly perturbed black holes in these tests always evolved into
non-singular time independent states, which suggests that the new
damped-wave gauge conditions introduced here may prove to be useful
for numerical simulations of more general dynamical black-hole
spacetimes as well.

We describe in some detail a number of technical properties of this
new gauge-driver system in a series of Appendices.  In
Appendix~\ref{s:Hyperbolicity} we show that any member of this new
class of first-order gauge drivers can be coupled to the GH Einstein
system in a way that makes the combined system symmetric hyperbolic.
In Appendix~\ref{s:DualCoordinateFrameEvolution} we develop a
dual-coordinate frame version of this new gauge-driver system, which
is needed to evolve black-hole binary systems for example.  In
Appendix~\ref{s:Constraints} we analyze the evolution of the
constraints in the new combined GH Einstein gauge-driver system.  We
show that the constraints and their evolution equations are the same
as those of the pure GH Einstein system, hence the constraint damping
properties of the original GH Einstein system are also unchanged.  In
Appendix~\ref{s:BoundaryConditions} we construct boundary conditions
for the gauge-driver system.  In most cases these boundary conditions
turn out to be the same as those used for the pure GH Einstein system,
but their representations in terms of the characteristic fields of the
gauge-driver system are different in some cases.  We also introduce a
new constraint-preserving boundary condition for the ``gauge''
components of the spacetime metric in the GH Einstein system.

%%%%%%%%%%%%%%%%%%%%%%%%%%%%%%%%%%%%%%%%%%%%%%%%%%%%%%%%%%%%%%%%%%%%%%%%%%%%%%%
\section{First-Order Gauge Driver}
%%%%%%%%%%%%%%%%%%%%%%%%%%%%%%%%%%%%%%%%%%%%%%%%%%%%%%%%%%%%%%%%%%%%%%%%%%%%%%%
\label{s:FirstOrderGaugeDriver}

The gauge drivers previously introduced for the GH Einstein
system~\cite{Pretorius2005a,Lindblom2007,Scheel2008} were constructed
by elevating the gauge-source function $H_a$ to the status of a
dynamical field that is evolved by a second-order wave
equation for $H_a$ having the general form,
\begin{eqnarray}
\psi^{cd}\partial_c\partial_d H_a = Q_a(H,\partial H,\psi,\partial\psi).
\label{e:GenericSecondOrderGaugeDriver}
\end{eqnarray}
When this type of evolution equation for $H_a$ is used together with
the GH Einstein evolution Eq.~(\ref{e:Einstein}), the combined system
is manifestly hyperbolic.  The first implementations of this type of
gauge driver were fairly successful, allowing a few successful binary
black-hole inspiral, merger and ringdown
simulations~\cite{Pretorius2005a,Scheel2008}.  A disadvantage of these
first gauge drivers however is that they were not designed to drive
$H_a$ toward a predetermined target $F_a$, so using them made it
difficult or impossible to predict what gauge would ultimately be
imposed on the solution.  One reason for this ambiguity is the
dynamical complexity of the operator used to evolve $H_a$.  Even the
homogeneous driver, Eq.~(\ref{e:GenericSecondOrderGaugeDriver}) with
$Q_a=0$, has a wealth of solutions that are not naturally attracted
toward any particular target $F_a$.  So it is not surprising that
these first gauge drivers have not been found to be very effective for
implementing pre-determined gauge conditions or for performing
evolutions in generic situations.  The goal here is to introduce a
gauge driver that drives $H_a$ toward a predetermined gauge specified
by $F_a$ more robustly and in more generic situations than was
possible with the first gauge drivers of this type~\cite{Lindblom2007}
based on the the complicated second-order wave operator used in
Eq.~(\ref{e:GenericSecondOrderGaugeDriver}).

An ideal gauge-driver would determine $H_a$ from an evolution
equation like,
\begin{eqnarray}
\partial_t H_a = - \mu(H_a-F_a),
\label{e:IdealGaugeDriver}
\end{eqnarray}
whose solutions all approach the target gauge-source function $F_a$
exponentially, at a rate determined by the freely specifiable
parameter $\mu$.  Unfortunately the evolution system formed by
combining Eq.~(\ref{e:IdealGaugeDriver}) with the GH Einstein
evolution Eq.~(\ref{e:Einstein}), does not appear to be hyperbolic.
There is a simple generalization of this ideal gauge driver however
that can be used with the GH Einstein equations to construct a
composite evolution system that is hyperbolic.  Let $t^a$ denote the
future-directed normal to the constant-$t$ hypersurfaces.  Then the
first-order gauge driver,
\begin{eqnarray}
t^b\partial_b H_a = - \tilde\mu (H_a-F_a),
\label{e:NaiveGaugeDriver}
\end{eqnarray}
combined with the GH Einstein evolution Eq.~(\ref{e:Einstein}) turns
out to be a hyperbolic system.  

We present a proof below that the combined GH Einstein gauge-driver
system, Eqs.~(\ref{e:Einstein}) and (\ref{e:NaiveGaugeDriver}), is
hyperbolic.  Before turning to that technical issue in
Appendix~\ref{s:Hyperbolicity} however, we point out that the very
simple gauge driver, Eq.~(\ref{e:NaiveGaugeDriver}), has some
limitations which can be overcome to some extent by a simple
modification.  To see these limitations we introduce spacetime
coordinates, $\{t,x^i\}$, where the time coordinate $t$ labels the
leaves in a foliation of spacelike hypersurfaces on which the points
are identified by the spatial coordinates $x^i$.  In this coordinate
system we use the standard 3+1 representation of the spacetime metric,
$\psi_{ab}$:
\begin{eqnarray}
ds^2&=&\psi_{ab}dx^adx^b,\nonumber\\
&=&-N^2 dt^2 + g_{ij}(dx^i+N^idt)(dx^j+N^jdt),
\label{e:ADMmetric}
\end{eqnarray}
where $g_{ij}$ is the intrinsic spatial metric of the constant-$t$
hypersurfaces, and $N$ and $N^i$ are referred to as the lapse and
shift respectively.  (We use Latin letters from the middle of the
alphabet, $i$, $j$, $k$, ..., for purely spatial indices.)  The unit
normal to the constant-$t$ hypersurfaces, $t^a$, has the 3+1
representation $t^a\partial_a =N^{-1}(\partial_t - N^k\partial_k)$
in this notation.  Thus the gauge driver given in
Eq.~(\ref{e:NaiveGaugeDriver}) can be written more explicitly in 3+1
form as
\begin{eqnarray}
\partial_t H_a -N^k\partial_k H_a = - \mu (H_a-F_a),
\end{eqnarray}
where $\mu= \tilde\mu N$.  This gauge driver has the property that
$H_a$ is driven toward $F_a$ as seen by observers moving along the
world lines of the hypersurface normal $t^a$.  However at a fixed
spatial coordinate, $x^i$, the quantity $H_a-F_a$ is not
necessarily driven to zero.  Therefore the evolution of a
dynamical spacetime (e.g., a perturbed black hole) using this driver
will not evolve toward a time independent state in which $H_a=F_a$.
Rather this driver will tend to evolve solutions into states with
$N^k\partial_k H_a = \mu (H_a-F_a)$.  This gauge may provide a
reasonable representation of the spacetime, but it will not be the
gauge $H_a=F_a$ the driver was intended to enforce.

This limitation in the gauge driver of Eq.~(\ref{e:NaiveGaugeDriver})
can be overcome by introducing an additional dynamical field,
$\theta_a$ defined as
\begin{eqnarray}
\partial_t \theta_a + \eta\,\theta_a = -\eta\, N^k\partial_k H_a.
\label{e:GaugeThetaEq}
\end{eqnarray}
or equivalently,
\begin{eqnarray}
\theta_a(t) = -\eta \int_{-\infty}^t e^{\eta(t'-t)}N^k\partial_k H_a(t') dt'.
\end{eqnarray}
The $\theta_a$ field is an exponentially weighted time average of
$-N^k\partial_k H_a$, which can be used to modify the gauge driver of
Eq.~(\ref{e:NaiveGaugeDriver})~\cite{Lindblom2007}:
\begin{eqnarray}
\partial_t H_a -N^k\partial_k H_a = - \mu (H_a-F_a) + \theta_a.
\label{e:GaugeHEq}
\end{eqnarray}
All time independent solutions of the first-order gauge driver
consisting of Eqs.~(\ref{e:GaugeThetaEq}) and (\ref{e:GaugeHEq}) must
now satisfy the desired gauge condition $H_a=F_a$.  Since the
gauge-driving parameters $\eta$ and $\mu$ are freely specifiable, they
can be chosen to enforce the desired gauge on a timescale shorter than
the characteristic time $\tau$ on which the spacetime evolves.  Thus
we expect the desired gauge can be enforced using this driver with
reasonable accuracy $H_a\approx F_a$ in any spacetime.  

In Sec.~\ref{s:NumericalTests} we present numerical tests of this
first-order gauge driver that demonstrate how well it succeeds.  In a
series of Appendices we also present some formal analyses of a variety
of mathematical properties of the new gauge driver composed of
Eqs.~(\ref{e:GaugeThetaEq}) and (\ref{e:GaugeHEq}) together with the
GH Einstein Eq.~(\ref{e:Einstein}).  In particular we show in
Appendix~\ref{s:Hyperbolicity} that this combined GH Einstein
gauge-driver system is symmetric hyperbolic.  In
Appendix~\ref{s:DualCoordinateFrameEvolution} we construct a
dual-coordinate frame version of this gauge driver that can be used
for example in the evolution of binary black-hole spacetimes.  In
Appendix~\ref{s:Constraints} we analyze the constraints and the
evolution of the constraints in the GH Einstein gauge-driver system.
And in Appendix~\ref{s:BoundaryConditions} we formulate boundary
conditions for the new gauge-driver system.

%%%%%%%%%%%%%%%%%%%%%%%%%%%%%%%%%%%%%%%%%%%%%%%%%%%%%%%%%%%%%%%%%%%%%%%%%%%%%%
\section{Numerical Tests}
\label{s:NumericalTests}
%%%%%%%%%%%%%%%%%%%%%%%%%%%%%%%%%%%%%%%%%%%%%%%%%%%%%%%%%%%%%%%%%%%%%%%%%%%%%%

In this section we describe the results of 3D numerical tests of the
 new GH Einstein gauge-driver system.  These tests evolve a
 Schwarzschild black hole with perturbed lapse and shift using the
 full coupled non-linear equations for the GH Einstein gauge-driver
 system, as described in Sec.~\ref{s:FirstOrderGaugeDriver}.  We
 measure the stability and effectiveness of the new gauge-driver
 system as it attempts to drive this single black-hole spacetime from
 the isotropic maximal-slicing gauge used to specify the initial data
 to an interesting new damped-wave gauge introduced in
 Appendix~\ref{s:DampedHarmonicGaugeConditions}.

These numerical tests are conducted using the infra\-structure of the
Caltech/Cornell Spectral Einstein Code (SpEC).  This code uses
pseudo-spectral collocation methods, as described for example in
Refs.~\cite{Kidder2005,Boyle2006}.  We use the generalized harmonic
form of the Einstein equations, as described in
Ref.~\cite{Lindblom2006}, together with the new gauge driver
Eqs.~(\ref{e:GaugeThetaEq}) and (\ref{e:GaugeHEq}).  Some of the tests
reported here use the dual-coordinate frame version of the new
gauge-driver system described in
Appendix~\ref{s:DualCoordinateFrameEvolution}.  For these dual-frame
tests we use the static Schwarzschild coordinates as the ``inertial''
frame, and a ``co-moving'' frame that rotates uniformly at angular
velocity $\Omega$ with respect to the inertial frame.  The evolution
equations for the combined GH Einstein gauge-driver system are
integrated in time using the method of lines and the adaptive
fifth-order Dormand-Prince integrator~\cite{DormandPrince1980}.

Initial conditions are needed for any evolution of the combined GH
Einstein gauge-driver system.  These initial data consist of the
spacetime metric $\psi_{ab}$, its time derivative $\partial_t
\psi_{ab}$, the gauge-source function $H_a$, and the time averaging
field $\theta_a$.  For the tests described here we take the initial
spacetime metric $\psi_{ab}$ to be the Schwarzschild geometry plus
perturbations as described below.  We set the time derivatives of the
spatial components of the metric initially to zero, and the
time-derivatives of the lapse and shift, $\partial_t N$ and
$\partial_t N^i$, are chosen to make $N$ and $N^i$ initially time
independent.  For the dual-frame evolution tests described below,
these time derivatives are chosen to make $N$ and the co-moving frame
components of $N^i$ time independent initially in the co-moving frame.
The initial value of $H_a$ is chosen to enforce the gauge constraint,
${\cal C}_a=H_a+\Gamma_a=0$, initially.  The value of the time
averaging field $\theta_a$ is set initially to ensure that its time
derivative vanishes, as determined by Eq.~(\ref{e:GaugeThetaEq}) or
(\ref{e:MovingGaugeTheta}).

For these tests we construct initial data consisting of a
Schwarzschild black hole with perturbations in the lapse and shift.
For the unperturbed hole we use isotropic spatial coordinates and
maximal time slices~\cite{Estabrook1973,Cook2004}.  The unperturbed
spatial metric in this representation is given by,
\begin{eqnarray}
ds^2&=&g_{ij}dx^idx^j
=\left(\frac{R}{r}\right)^2\left(dx^2
+ dy^2 + dz^2\right),\label{e:IsotropicMetric}\qquad
\end{eqnarray}
where $r^2=x^2+y^2+z^2$, and $R(r)$ (the areal radius)
satisfies the differential equation,
\begin{eqnarray}
\frac{{\mathrm d}R}{{\mathrm d}r} &=&
\frac{R}{r}\sqrt{1-\frac{2M}{R}+\frac{C^2}{R^4}}.
\end{eqnarray}
The constant $M$ is the mass of the hole, and $C$ is a parameter that
specifies the particular maximal slicing.  Finally, the unperturbed
lapse $N$ and shift $N^i$ for this representation of Schwarzschild are
given by,
\begin{eqnarray}
N&=& \sqrt{1-\frac{2M}{R}+\frac{C^2}{R^4}},\label{e:SchwarzschildLapse}\\
N^i&=&\frac{C\hat r^i}{R^2}
\left(1-\frac{2M}{R}+\frac{C^2}{R^4}\right),\label{e:SchwarzschildShift}
\end{eqnarray}
where $\hat r^i$ is the outward directed radial unit vector:
$g_{ij}\hat r^i \hat r^j=1$.

We perturb this spacetime by changing the initial values of the lapse
and shift, and their time derivatives.  This type of perturbation
changes the spacetime coordinates (or gauge) of the solution, but not
its geometry.  For these tests we modify the lapse and shift of
Eqs.~(\ref{e:SchwarzschildLapse}) and (\ref{e:SchwarzschildShift}) by
adding perturbations of the form,
\begin{eqnarray}
\delta N &=& A \sin(2\pi r/r_0)e^{-(r-r_c)^2/w^2}Y_{lm},
\label{e:PerturbedSchwarzschildLapse}\\
\delta N^i &=& A \sin(2\pi r/r_0)e^{-(r-r_c)^2/w^2} Y_{lm}\hat r^i,
\label{e:PerturbedSchwarzschildShift}
\end{eqnarray}
where $Y_{lm}$ is the standard scalar spherical harmonic.  In our
numerical tests we use the background metric with $C=1.73M^2$, and
perturbations with $A=0.01$, $r_c=15M$, $w=3M$, 
$r_0=6M$, and $l=2$, $m=0$.

These numerical tests are performed using the target gauge-source function
for the new damped-wave gauge,
\begin{equation}
F_a=\mu_L\log\left(\frac{g^{\,p}}{N}\right)t_a-\mu_S N^{-1}g_{ai}N^i,
\label{e:FullDampedWaveGaugeText}
\end{equation}
where $\mu_L$ and $\mu_S$ are damping parameters, $g=\det g_{ij}$, and
$p$ is a constant.  This new gauge condition is discussed in some
detail in Appendix~\ref{s:DampedHarmonicGaugeConditions}.  The gauge
used to prepare the perturbed Schwarzschild initial data,
Eqs.~(\ref{e:IsotropicMetric})--(\ref{e:PerturbedSchwarzschildShift}),
is very different from the damped-wave gauge condition.  It is always
difficult to start evolutions in a smooth and convergent way using
initial data prepared with a significantly different gauge.  To
minimize this start-up problem, it is common practice to turn on the
new gauge condition gradually.  We do this in our gauge driver system
by defining an initial target $F_a^{(0)}$ that is simply the
constraint-satisfying $H_a$ of the unperturbed initial data.  Except
for the perturbation, this is exactly the gauge needed for a time
independent evolution of these initial data.  We then set the target
$F_a$ to
\begin{equation}
F_a = e^{-t^2/T^2} F_a^{(0)}
    + \left(1-e^{-t^2/T^2}\right) F^{DW}_a,
\end{equation}
where $F^{DW}_a$ is the target gauge-source function for the
damped-wave gauge defined in Eq.~(\ref{e:FullDampedWaveGaugeText}).  This
choice for $F_a$ changes the gauge condition from its initial state
$F_a^{(0)}$ to the desired $F^{DW}_a$ smoothly and gradually on the
timescale $T$.  For the tests discussed here we use $T=10 M$ for the
value of this time-blending parameter.  

These tests use the damped-wave gauge condition defined in
Eq.~(\ref{e:FullDampedWaveGaugeText}) with damping parameters
$\mu_S=\mu_L=0.1$ and $p=0.5$.  Most of these tests (except as noted
below) use the values $\mu=\eta=16$ for the gauge-driver parameters,
used in Eqs.~(\ref{e:GaugeThetaEq}) and (\ref{e:GaugeHEq}), and the
boundary gauge-driver parameter $\mu_B=1$ used in
Eq.~(\ref{e:HDriverBC}).  These tests set the constraint damping
parameters of the GH Einstein system to the values:
$\gamma_0=\gamma_2=2$ and $\gamma_1=-1$, cf. Ref.\cite{Lindblom2006}.

We perform these numerical tests on a computational domain consisting
of a spherical shell that extends from $r=0.78M$ (just inside the
horizon in the initial coordinates) to $r=60M$ (well outside the
domain of influence of the initial perturbations).  We divide this
domain into sixteen sub-domains, which allows us to distribute the
computation over several processors to enhance computational speed.
In each sub-domain we express each Cartesian component of each
dynamical field as a sum of Chebyshev polynomials of $r$ (through
order $N_r-1$) multiplied by scalar spherical harmonics (through order
$L$).  The radii of the inner and outer edges of the various
sub-domains are adjusted to distribute the truncation error during the
full time evolution more or less uniformly on the grid.  The specific
radii of the sub-domain boundaries used in these tests are $0.78M,$
$1.68M,$ and $k\times 4.0M$ for $k=1,\, ...\, ,15$.

In the pseudo-spectral numerical method used here, each
Cartesian component of each dynamical field is expanded as a sum of
the form:
\begin{eqnarray}
u(r,\theta,\varphi)
=\sum_{k=0}^{N_r-1}\sum_{\ell=0}^{L}\sum_{m=-\ell}^{\ell} u_{k\ell m}
T_k(r)Y_{\ell m}(\theta,\varphi),
\end{eqnarray}
where the $u_{k\ell m}$ are referred to as the spectral coefficients
of the field $u$.  These spectral coefficients must be modified in
this method through a process called spectral filtering.  We use two
types of spectral filtering in these tests.  One type affects the
angular spectral coefficients, as described in Ref.~\cite{Kidder2005}.
This filter sets to zero in each time step the changes in the top four
{\it tensor} spherical harmonic expansion coefficients of each of the
dynamical fields.  This filtering step is needed to eliminate an
instability associated with the inconsistent mixing of tensor
spherical harmonics whenever angular derivatives are computed in our
approach.  In addition we also perform the following radial filtering,
\begin{equation}
{\cal F}(u_{k\ell m})=e^{-[ k/\rho\,(N_r-1)]^p} u_{k\ell m},
\end{equation}
where ${\cal F}(u_{k\ell m})$ represents the filtered coefficients,
before applying outer boundary conditions as described in
Appendix~\ref{s:BoundaryConditions}.  For these tests we use
$\rho=0.9$ and $p=18$, which leaves essentially unchanged the
coefficients $u_{k\ell m}$ with $k \lesssim 2 (N_r-1)/3$, while the
coefficient of the highest mode, $k=N_r-1$, is effectively set to zero.
This radial filter implements in a smooth way the standard $2/3$
filter often used to cure non-linear aliasing that can occur in
spectral evolutions~\cite{Boyd1999,Gottlieb2001}.

\begin{figure}[t]
\vspace{0.1cm}
\centerline{
\includegraphics[width=3in]{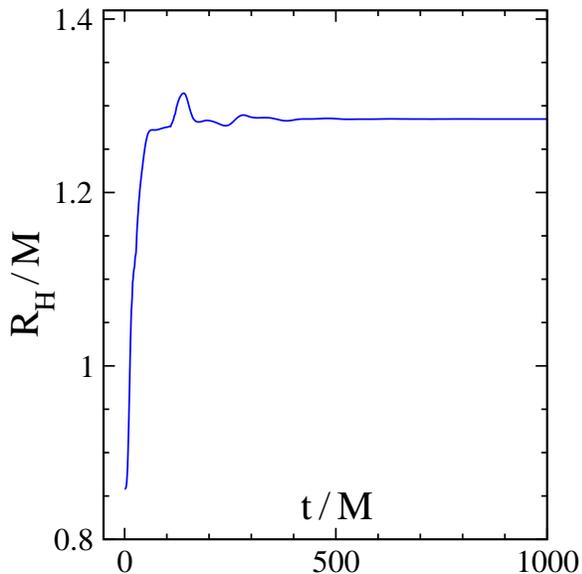}}
\caption{\label{f:BHRadiusHorizon}Coordinate radius of the
apparent horizon of the black hole $R_H$ as it evolves under the
effects of the dynamically driven gauge.  This test uses a
single-frame evolution with gauge-driver parameters $\mu=\eta=16$.}
\end{figure}

The damped-wave gauge conditions defined by
Eq.~(\ref{e:FullDampedWaveGaugeText}) (and described in
Appendix~\ref{s:DampedHarmonicGaugeConditions}) are significantly
different than those satisfied by the perturbed maximally sliced
representation of the Schwarzschild geometry used as initial data for
this test.  Consequently the representation of the black hole in our
test becomes very dynamical, primarily due to these gauge differences,
and also due to the presence of the asymmetric perturbation applied to
the lapse and shift.  Figure~\ref{f:BHRadiusHorizon} illustrates just
how significant these gauge differences are by showing the evolution
of the coordinate radius of the apparent horizon $R_H$ of the black hole.  In
these tests the radius of the apparent horizon $R_H$ grows by 50\%,
changing from an initial value of $0.86M$ to a final radius of
$1.28M$.
%
%%%%%%%%%%%%%%%%%%%%%%%%%%%%%%%%%%%%%%%%%%%%%%%%%%%%%%%%%%%%%%%%%%%%%%%%%%%%%%%
%%%%%%%%%%%%%%%%%%%%%%%%%%%%%%%%%%%%%%%%%%%%%%%%%%%%%%%%%%%%%%%%%%%%%%%%%%%%%%%

\begin{figure}[t]
\vspace{0.1cm}
\centerline{
\includegraphics[width=3in]{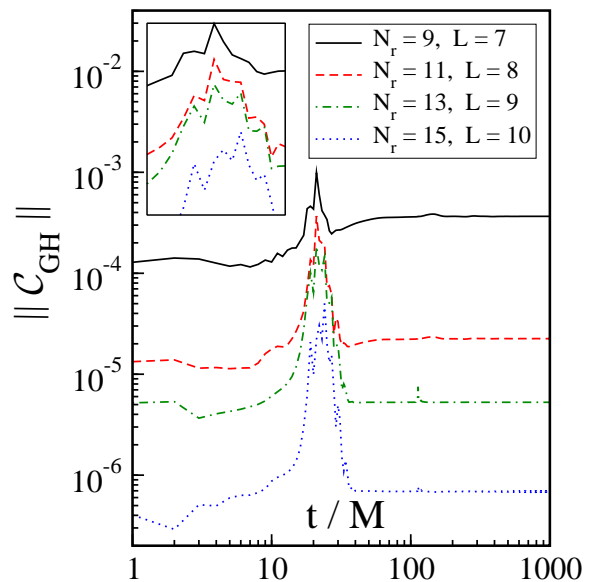}}
\caption{\label{f:BHGaugeConstraints} Constraints of the GH Einstein
  system $||\,{\cal C}_{\mathrm{GH}}||$ for a single-frame evolution
  of a Schwarzschild black hole with strongly perturbed lapse and
  shift.  This test uses a single-frame evolution with gauge-driver
  parameters $\mu=\eta=16$, and several different values of the
  numerical resolution parameters $N_r$ and $L$.  The small inset
  graph contains a magnified view of $||\,{\cal C}_{\mathrm{GH}}||$
  during the time interval $15M\leq t \leq 30M$, showing that the
  solution is convergent during this most dynamical part of the
  evolution.}
\end{figure}

\begin{figure}[t]
\vspace{0.1cm}
\centerline{
\includegraphics[width=3in]{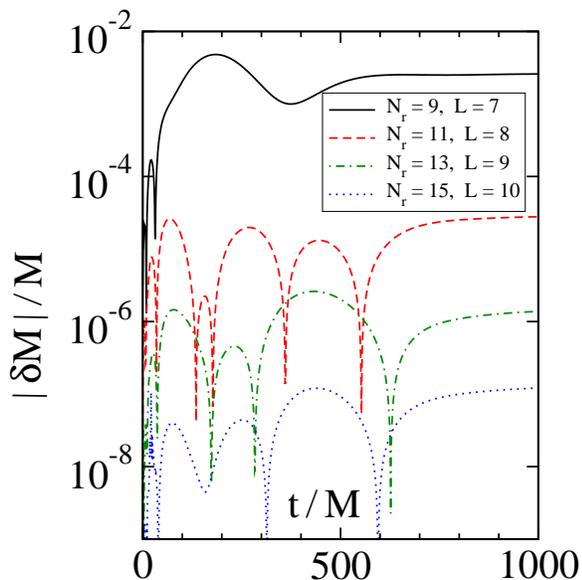}}
\caption{\label{f:BHMassVariations}Curves show $|\delta M|/M$,
the deviations in the mass of the hole from its initial value.
This test uses a single-frame evolution with gauge-driver parameters
$\mu=\eta=16$.}
\end{figure}

Figure~\ref{f:BHGaugeConstraints} illustrates the constraint
violations for a single-frame evolution of the GH Einstein
gauge-driver system, and demonstrates the stability and convergence of
our numerical method.  The constraints of the GH Einstein gauge-driver
system are identical to those of the GH Einstein system, as discussed
in some detail in Appendix~\ref{s:Constraints}.  Therefore we measure
constraint violations using the quantity $||\,{\cal
  C}_{{\,\mathrm{GH}}}||$, the ratio of an $L^2$ norm of all the GH
Einstein constraints divided by an $L^2$ norm of the derivatives of
the dynamical fields.  This constraint norm vanishes iff the
constraints are satisfied, and has been normalized to be of order
unity when constraint violations begin to dominate the solution.  This
constraint norm was originally introduced to measure constraint
violations for the pure GH Einstein system in Eq.~(71) of
Ref.~\cite{Lindblom2006}. The constraint violations become largest and
the rate of convergence of the simulations decreases during the time
interval $15M\lesssim t \lesssim 30M$ in
Fig.~\ref{f:BHGaugeConstraints} when the inward moving gauge
perturbation interacts most strongly with the black hole.  These
results show that the constraints are well satisfied throughout the
evolutions, demonstrates that our numerical methods are convergent,
and shows that the GH Einstein gauge-driver system is stable over many
dynamical timescales.  Figure~\ref{f:BHMassVariations} provides
another illustration of the stability and the numerical convergence of
the GH Einstein gauge-driver system.  In this figure we show $|\delta
M(t)|/M$ the evolution of the difference between the evolved and the
initial mass of the black hole (as determined from the area of its
apparent horizon).
%
%%%%%%%%%%%%%%%%%%%%%%%%%%%%%%%%%%%%%%%%%%%%%%%%%%%%%%%%%%%%%%%%%%%%%%%%%%%%%%%
%%%%%%%%%%%%%%%%%%%%%%%%%%%%%%%%%%%%%%%%%%%%%%%%%%%%%%%%%%%%%%%%%%%%%%%%%%%%%%%
%%%%%%%%%%%%%%%%%%%%%%%%%%%%%%%%%%%%%%%%%%%%%%%%%%%%%%%%%%%%%%%%%%%%%%%%%%%%%%%
%%%%%%%%%%%%%%%%%%%%%%%%%%%%%%%%%%%%%%%%%%%%%%%%%%%%%%%%%%%%%%%%%%%%%%%%%%%%%%%
%%%%%%%%%%%%%%%%%%%%%%%%%%%%%%%%%%%%%%%%%%%%%%%%%%%%%%%%%%%%%%%%%%%%%%%%%%%%%%
\begin{figure}[t]
\vspace{0.1cm}
\centerline{
\includegraphics[width=3in]{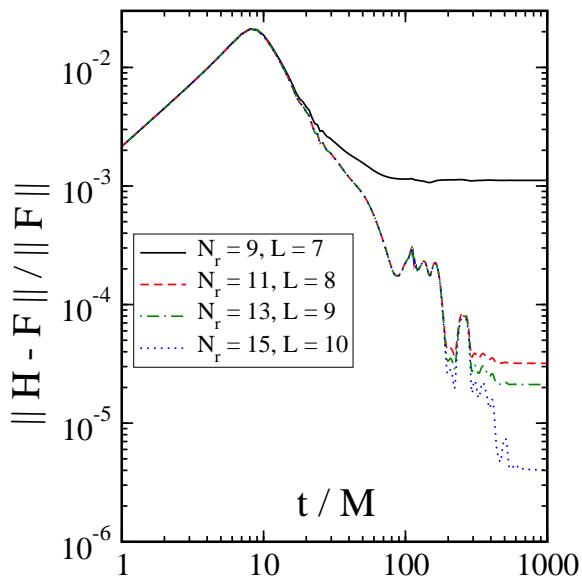}}
\caption{\label{f:HMinusFOmega0}Effectiveness of the gauge-driver
equation is demonstrated by showing $||H-F||/||F||$ for an evolution
of a Schwarzschild black hole with strongly perturbed lapse and shift.
This test uses a single-frame evolution with gauge-driver parameters
$\mu=\eta=16$. }
\end{figure}
%%%%%%%%%%%%%%%%%%%%%%%%%%%%%%%%%%%%%%%%%%%%%%%%%%%%%%%%%%%%%%%%%%%%%%%%%%%%%%%
%

\begin{figure}[t]
\vspace{0.1cm}
\centerline{
\includegraphics[width=3in]{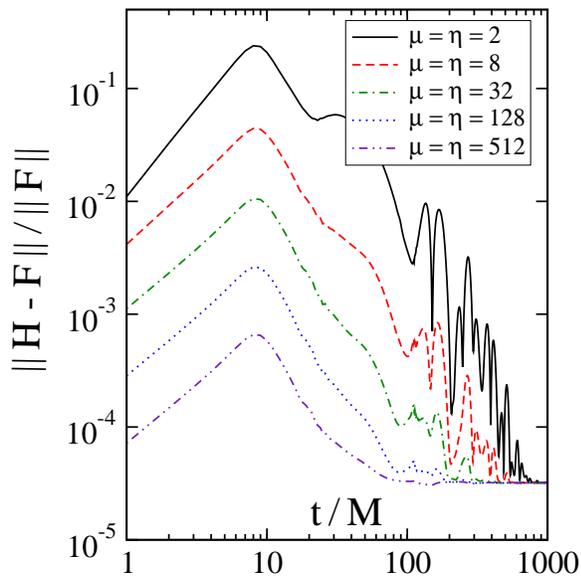}}
\caption{\label{f:HMinusFVaryingMu}Effectiveness of the gauge-driver
  system is demonstrated for various values of the gauge-driver
  parameters $\eta$ and $\mu$.  This test uses a dual-frame evolution
  method with the co-moving frame rotating with respect to the
  inertial frame at angular velocity $\Omega=1/M$.}
\end{figure}

Figure~\ref{f:HMinusFOmega0} demonstrates the effectiveness of the
gauge-driver system for this test problem.  The difference between the
gauge source function $H_a$ and the target function to which it is
being driven, $F_a$, is measured using the following $L^2$ norm:
\begin{eqnarray}
\frac{||H-F||^2}{||F||^2}&=&
\frac{\int\!\sqrt{g}\,m^{ab}(H_a-F_a)(H_b-F_b)\,d^{\,3}x}
{\int\!\sqrt{g}\, m^{cd}F_cF_d\,d^{\,3}x},\qquad
\end{eqnarray}
where $m^{ab}$ is a positive definite matrix, set to the identity,
$m^{ab}=\delta^{ab}$, for these tests.  This norm vanishes if and only
if the target gauge condition, $H_a=F_a$ is satisfied, and it is
scaled so that $H_a$ bears little resemblance to the target $F_a$
whenever it becomes of order unity.  Figure~\ref{f:HMinusFVaryingMu}
shows that the initial mismatch between the gauge of the perturbed
black hole and the damped wave gauge conditions (defined by $F_a$)
causes $||H-F||/||F||$ to grow initially.  But the gauge-driver steps
in and limits this growth to a maximum of about 0.02 in these
evolutions, and then drives $||H-F||/||F||$ to very small values
(depending on the numerical resolution) at late times.

The evolution tests illustrated in
Figs.~\ref{f:BHRadiusHorizon}--~\ref{f:HMinusFOmega0} were performed
using the single-frame version of the gauge-driver system described in
Sec.~\ref{s:FirstOrderGaugeDriver}.  Binary black-hole simulations are
done with the Caltech/Cornell SpEC code using a dual-coordinate frame
formulation of the GH Einstein equations~\cite{Scheel2006}.  In this
formulation the components of the various tensor fields are defined
with respect to a non-rotating inertial coordinate frame, while the
equations for these field components are solved using a co-moving
coordinate frame that tracks the motions of the black holes.  A
dual-frame version of the GH Einstein gauge-driver system is developed
in Appendix~\ref{s:DualCoordinateFrameEvolution}.  We have performed
the same perturbed single black-hole evolution tests illustrated in
Figs.~\ref{f:BHRadiusHorizon}--~\ref{f:HMinusFOmega0}. using this
dual-frame version of the GH Einstein gauge-driver system.  For these
tests we use a co-moving frame that rotates with respect to the
asymptotic inertial frame at angular velocity $\Omega=1/M$.  (This
means that equatorial grid points in this test move at 60 times the
speed of light at the outer edge of our computational domain.)  The
gauge driver used for these evolutions is the hybrid driver described
in Appendix~\ref{s:DualCoordinateFrameEvolution},
Eqs.~(\ref{e:HybridDualGaugeHEq}) and
(\ref{e:HybridDualGaugeThetaEq}).  This driver attempts to enforce the
comoving-frame gauge condition $H_a=F_a$ in the spacetime region near
the black hole, while enforcing the inertial-frame condition $H_{\bar
  a}=F_{\bar a}$ near the outer boundary of the computational domain.
The transition between these is accomplished by smoothly
blending the two conditions at intermediate points using a weight
function $w(x)$, cf. Eqs.~(\ref{e:HybridDualGaugeHEq}) and
(\ref{e:HybridDualGaugeThetaEq}).  In regions where $w(x)=1$, the pure
comoving-frame condition is enforced, and where $w(x)=0$ the pure
inertial-frame condition is used.  For these numerical tests we use
$w(r)=e^{-[r/(0.89\,R_o)]^{17}}$, where $R_o=60M$ is the outer radius
of the computational domain.  This choice accurately enforces the
comoving-frame condition in the inner region of the domain where
$r\lesssim 2R_o/3$, and the inertial-frame condition at points 
located very near the outer boundary, $r\approx R_o$.

The graphs of the quantities depicted in
Figs.~\ref{f:BHRadiusHorizon}--~\ref{f:HMinusFOmega0} for the
dual-frame evolution case are almost identical to their single-frame
evolution counterparts.  So we will not show those graphs again here.
Instead we show in Fig.~\ref{f:HMinusFVaryingMu} a series of
evolutions performed with the dual-frame system in which the effects
of varying the gauge-driver parameters $\mu$ and $\eta$ are examined.
We see from these results, that the gauge-driver system is very
effective in driving $H_a\rightarrow F_a$ for a wide range of
gauge-driver parameters.  Evolutions using larger values of the
gauge-driver parameters are generally more effective in keeping the
quantity $||H-F||/||F||$ small and driving it quickly toward zero.
The gauge-driver system is stable and effective over a rather wide
range of parameters, but becomes ineffective when the gauge-driver
parameters get smaller than about one, and the system also becomes
unstable when the parameters are larger than a few hundred.
%
%%%%%%%%%%%%%%%%%%%%%%%%%%%%%%%%%%%%%%%%%%%%%%%%%%%%%%%%%%%%%%%%%%%%%%%%%%%%%%%
%%%%%%%%%%%%%%%%%%%%%%%%%%%%%%%%%%%%%%%%%%%%%%%%%%%%%%%%%%%%%%%%%%%%%%%%%%%%%%%
%

%%%%%%%%%%%%%%%%%%%%%%%%%%%%%%%%%%%%%%%%%%%%%%%%%%%%%%%%%%%%%%%%%%%%%%%%%%%%%%%
\appendix
%%%%%%%%%%%%%%%%%%%%%%%%%%%%%%%%%%%%%%%%%%%%%%%%%%%%%%%%%%%%%%%%%%%%%%%%%%%%%%%

%%%%%%%%%%%%%%%%%%%%%%%%%%%%%%%%%%%%%%%%%%%%%%%%%%%%%%%%%%%%%%%%%%%%%%%%%%%%%%
\section{Damped-Wave Gauge Conditions}
\label{s:DampedHarmonicGaugeConditions}
%%%%%%%%%%%%%%%%%%%%%%%%%%%%%%%%%%%%%%%%%%%%%%%%%%%%%%%%%%%%%%%%%%%%%%%%%%%%%%

Harmonic gauge is defined by the condition that each coordinate $x^a$
satisfies the co-variant scalar wave equation:
\begin{eqnarray}
\nabla^c\nabla_c x^a = H^a = 0.
\label{e:HarmonicGauge}
\end{eqnarray}
Harmonic coordinates have proven to be extremely useful for analytical
studies of the Einstein equations, but have found only limited success
in numerical problems like simulations of complicated highly dynamical
black-hole mergers.  A likely reason for some of these difficulties is
the wealth of ``interesting'' dynamical solutions to the harmonic
gauge condition itself, Eq.~(\ref{e:HarmonicGauge}).  Since all
``physical'' dynamical fields are expressed in terms of the
coordinates, an ideal gauge condition would limit coordinates to those
that are simple, straightforward, dependable, and non-singular; having
``interesting'' dynamics of their own is {\it not} a desirable feature
for coordinates.  We propose to reduce the dynamical range available
to harmonic coordinates by adding a damping term to the equation:
\begin{eqnarray}
\nabla^c\nabla_c x^a = \mu_S t^c\partial_c x^a = \mu_S t^a,
\end{eqnarray}
where $t^a$ is the future directed unit normal to the constant-$t$
hypersurfaces.  Adding such a damping term to the equations for the
spatial coordinates $x^i$ tends to remove extraneous gauge dynamics
and drives the coordinates toward solutions of the co-variant spatial
Laplace equation on the timescale $1/\mu$.  Choosing $1/\mu$ to be
comparable to (or smaller than) the characteristic timescale of a
particular problem should remove any extraneous coordinate dynamics on
timescales shorter than the physical timescale. The addition of such a
damping term in the time-coordinate equation is not appropriate
however.  Such a damped-wave time coordinate is driven toward a
constant value, and therefore toward a state in which it fails to be a
useful time coordinate at all. It makes sense then to use the
damped-wave gauge condition only for the spatial coordinates:
\begin{eqnarray}
\nabla^c\nabla_c x^i = H^i = \mu_S t^i = - \mu_S N^i/N,
\label{e:DampedWaveShift}
\end{eqnarray}
where $N^i$ is the shift, and $N$ is the lapse.  The appropriate
contra-variant version of this damped-wave gauge condition is
therefore
\begin{eqnarray}
H_a = -\mu_S g_{ai} N^i / N,
\label{e:DampedWaveCoordinates}
\end{eqnarray}
where $g_{ab}=\psi_{ab}+t_a t_b$ is the spatial
  metric.\footnote{Frans Pretorius and Matthew Choptuik have recently, 
  independently,
  proposed adding similar damping terms to the harmonic gauge
  condition~\cite{Choptuik09}.}

  We point out that the damped-wave gauge condition,
  Eq.~(\ref{e:DampedWaveCoordinates}), is very similar to one version
  of the $\Gamma$-driver shift condition adopted recently by several
  groups using moving puncture evolution methods~\cite{vanMeter2006}.
  It is straightforward to express the co-variant wave operator in
  terms of the 3+1 decomposition of the metric:
\begin{eqnarray}
\nabla^c\nabla_c x^i &=& - {}^{(3)}\Gamma^i+
N^{-2}\left(\partial_t N^i - N^k\partial_k N^i\right)\nonumber\\
&&\qquad+ g^{ik}\partial_k \log N  ,
\end{eqnarray}
where ${}^{(3)}\Gamma^i$ is the trace of the Christoffel connection
computed from $g_{ij}$.  It follows that the damped-wave
shift condition, Eq.~(\ref{e:DampedWaveShift}),
is equivalent to the following condition on the shift:
\begin{eqnarray}
&&\partial_t N^i - N^k\partial_k N^i+\mu_S N N^i =\nonumber\\
&&\qquad\qquad\qquad N^{2}\left[
{}^{(3)}\Gamma^i- g^{ik}\partial_k \log N\right].
\end{eqnarray}
In comparison a version of the $\Gamma$-driver shift condition,
cf. Eq. (26) of Ref.~\cite{vanMeter2006}, that is currently
being used by a number of numerical relativity groups
is a very similar condition:
\begin{eqnarray}
\partial_t N^i - N^k\partial_k N^i+\eta N^i =0.75\,
{}^{(3)}\tilde\Gamma^i,
\end{eqnarray}
where ${}^{(3)}\tilde\Gamma^i$ is the trace of the Christoffel
connection computed from the conformal metric $\tilde g_{ij}= g^{-1/3}
g_{ij}$.  This version of the $\Gamma$-driver shift condition is
therefore a certain conformal damped-wave equation for the spatial
coordinates.

While the damped-wave gauge is a poor choice for the time
coordinate, the idea of imposing a gauge that uses the dissipative
properties of the damped-wave equation to suppress extraneous
gauge dynamics is attractive.  The lapse is the rate of change of
proper time with respect to the time coordinate (as measured by an
observer moving along $t^a$), so choosing a gauge in which the lapse
satisfies a damped-wave equation seems like the appropriate
time-domain analog of the damped-wave spatial gauge condition. To
find the appropriate expression for $t^aH_a$ that leads to such an
equation, we note that the gauge constraint $H_a+\Gamma_a=0$ implies
that $t^aH_a$ is given by
\begin{eqnarray}
t^a H_a = -K - t^a\partial_a \log N,
\label{e:BasicLapseConstraint}
\end{eqnarray}
where $K=g^{ij}K_{ij}$ is the trace of the extrinsic curvature of the
constant-$t$ hypersurfaces.  Using the definition of $K$, this
condition can be also be written in the form,
\begin{eqnarray}
t^a H_a =t^a\partial_a \log \left(\frac{\sqrt{g}}{N}\right)-
N^{-1}\partial_k N^k,
\label{e:AlternateLapseConstraint}
\end{eqnarray}
where $g=\det g_{ij}$ is the spatial
volume element.  One frequent symptom of the failure of simpler
gauge conditions in binary black-hole simulations is an explosive growth
in $g$ in the spacetime region near the black-hole horizons.  This
suggests choosing the gauge condition,
\begin{eqnarray}
t^a H_a =-\mu_L\log \left(\frac{\sqrt{g}}{N}\right)
\label{e:DampedWaveLapse}
\end{eqnarray}
for $\mu_L>0$, which tends to suppress any growth in $\sqrt{g}/N$ as a
consequence of the constraint, Eq.~(\ref{e:AlternateLapseConstraint}).

To determine how this gauge condition, Eq.~(\ref{e:DampedWaveLapse}),
effects the evolution of the lapse, we note that the time derivative
of $K$ is determined by the Einstein evolution equations:
\begin{eqnarray}
t^a\partial_a K = K_{ij}K^{ij} - N^{-1}D^iD_i N, 
\end{eqnarray}
where $D_i$ is the spatial co-variant derivative compatible with
$g_{ij}$.  Combining this expression with
Eq.~(\ref{e:BasicLapseConstraint}) gives an equation for the time
derivative of $t^aH_a$,
\begin{eqnarray}
N\, t^b\partial_b\bigl(t^aH_a\bigr)&=& 
-t^b\partial_b\bigl(t^a\partial_a N\bigr)
+D^iD_i N\nonumber\\
&& + N^{-1}(t^a\partial_a N)^2 - N K_{ij}K^{ij},
\label{e:LapseWaveEquation}
\end{eqnarray}
which is a wave operator acting on the lapse.  When the gauge
condition in Eq.~(\ref{e:DampedWaveLapse}) is enforced, it equates
this wave operator to the following expression,
\begin{eqnarray}
N\,t^b\partial_b\bigl(t^aH_a)= \mu_L t^a\partial_a N-{\scriptstyle \frac{1}{2}}
\mu_L N t^a\partial_a \log g.
\label{e:LapseDampingTerms}
\end{eqnarray}
The first term on the right side of Eq.~(\ref{e:LapseDampingTerms}) is
a standard damping term for the lapse wave equation, while the second
term plays the role of an additional ``source.''  The
motivation for including the particular dependence on $g$ in
Eq.~(\ref{e:LapseDampingTerms}) is provided by the argument leading to
Eq.~(\ref{e:DampedWaveLapse}), however, this dependence can easily be
generalized without changing the term's fundamental lapse-damping property 
by setting
\begin{eqnarray}
t^aH_a = -\mu_L\log\left(\frac{g^{\,p}}{N}\right),
\label{e:NewSlicingCondition}
\end{eqnarray}
where $p$ is a constant.  The case $p=0.5$ corresponds to
Eq.~(\ref{e:DampedWaveLapse}), while $p=0$ is a pure lapse-damping
gauge without the extra source term.  

Combining this new lapse condition, Eq.~(\ref{e:NewSlicingCondition}),
with the damped-wave spatial coordinate condition,
Eq.~(\ref{e:DampedWaveCoordinates}), gives the target gauge-source
function for our full damped-wave gauge condition:
\begin{eqnarray}
F_a=\mu_L\log\left(\frac{g^{\,p}}{N}\right)t_a-\mu_S N^{-1}g_{ai}N^i.
\label{e:FullDampedWaveGauge}
\end{eqnarray}
This gauge condition depends only on the spacetime metric $\psi_{ab}$,
so it could be implemented directly in the GH Einstein system by
setting $H_a=F_a$.  However it can also be implemented with the new GH
Einstein gauge-driver system introduced in
Sec.~\ref{s:FirstOrderGaugeDriver}, where it can be used as a
non-trivial test of the new gauge-driver.  Numerical evolutions of
strongly perturbed single black-hole spacetimes using the $p=0.5$
version of this gauge and the new GH Einstein gauge-driver system are
described in Sec.~\ref{s:NumericalTests}.

%%%%%%%%%%%%%%%%%%%%%%%%%%%%%%%%%%%%%%%%%%%%%%%%%%%%%%%%%%%%%%%%%%%%%%%%%%%%%%%
\section{Hyperbolicity}
\label{s:Hyperbolicity}
%%%%%%%%%%%%%%%%%%%%%%%%%%%%%%%%%%%%%%%%%%%%%%%%%%%%%%%%%%%%%%%%%%%%%%%%%%%%%%%

The hyperbolicity of an evolution system consisting of some
first-order equations, like our new gauge driver
Eqs.~(\ref{e:GaugeThetaEq}) and (\ref{e:GaugeHEq}), and some
second-order equations, like the GH Einstein Eq.~(\ref{e:Einstein}),
is most easily analyzed by converting all the equations to first-order
form.  The spectral evolution code that we use to perform our
numerical simulations is rather sensitive to ill-posed evolution
problems.  So we generally perform our numerical simulations by
evolving first-order systems of equations where hyperbolicity is
easier to analyze and where boundary conditions are easier to
construct.  Mixed systems like the combined Einstein and gauge-driver
equations can be converted to first-order form by introducing
additional dynamical fields for the first derivatives of those fields
satisfying second-order equations.  Convenient choices of the needed
additional fields for the GH Einstein system are
$\Pi_{ab}=-t^c\partial_c\psi_{ab}$ and
$\Phi_{iab}=\partial_i\psi_{ab}$.  The evolution equations for these
fields, $\{\psi_{ab},\Pi_{ab},\Phi_{iab}\}$, then become a first-order
representation of the GH Einstein system:
\begin{eqnarray}
%PsiDot
&&\!\!\!\!\!\!\!\!\!\!
\partial_t\psi_{ab}-(1+\gamma_1)N^k\partial_k\psi_{ab} 
 = - N\Pi_{ab}-\gamma_1N^i\Phi_{iab},\qquad
\label{e:NewPsiDot}\\
%PiDot
&&\!\!\!\!\!\!\!\!\!\!
\partial_t\Pi_{ab} - N^k\partial_k\Pi_{ab} 
+ N g^{ki}\partial_k\Phi_{iab} - 
\gamma_1 \gamma_2 N^k \partial_k \psi_{ab}
\nonumber\\
&&\!\!\!
+2N\partial_{(a}H_{b)}
=
- \half Nt^c t^d \Pi_{cd}\Pi_{ab}
-N t^c \Pi_{c i} g^{ij}\Phi_{jab}\nonumber\\
&&\,\,\,
+2N\psi^{cd}\bigl(  
  g^{ij} \Phi_{ica} \Phi_{jdb}
- \Pi_{ca} \Pi_{db}
- \psi^{ef}\Gamma_{ace}\Gamma_{bdf}
\bigr)
\nonumber\\
&&\,\,\,+N\gamma_0 \bigl[2\delta^c{}_{(a}t{}_{b)}-\psi_{ab}
t^c\bigr] ({H}_c+\psi^{ef}\Gamma_{cef})
\nonumber\\
&&\,\,\,+2N\Gamma^c_{ab}H_{c}
- \gamma_1 \gamma_2 N^i \Phi_{iab},\label{e:NewPiDot}\\
%PhiDot
&&\!\!\!\!\!\!\!\!\!\!
\partial_t\Phi_{iab}-N^k\partial_k\Phi_{iab}
+N\partial_i\Pi_{ab}-N\gamma_2\partial_i\psi_{ab}
\nonumber\\
&&\!\!\!=\half N t^c t^d \Phi_{icd}\Pi_{ab}
+Ng^{jk}t^c\Phi_{ijc}\Phi_{kab}
-N\gamma_2\Phi_{iab},\label{e:NewPhiDot}
\end{eqnarray}
cf. Eqs. (35)-(37) of Ref.~\cite{Lindblom2006}.  In these equations
$N$, $N^i$, and $g_{ij}$, are the standard 3+1 representation of
$\psi_{ab}$ given in Eq.~(\ref{e:ADMmetric}); $t^a$ is the future
directed timelike unit normal; $\Gamma^c_{ab}$ is the Christoffel
connection associated with $\psi_{ab}$; and $\gamma_0$, $\gamma_1$,
and $\gamma_2$ are parameters multiplying constraints, introduced
because they help damp away small constraint violations.  This
representation of the GH Einstein equations together with the gauge
driver introduced above, Eqs.~(\ref{e:GaugeThetaEq}) and
(\ref{e:GaugeHEq}), is a first-order evolution system which can be
represented abstractly as,
\begin{eqnarray}
\partial_t u^\alpha + A^{k\,\alpha}{}_\beta \partial_k u^\beta
= B^\alpha.
\label{e:FOSystem}
\end{eqnarray}
For the combined GH Einstein gauge-driver system, the collection of
dynamical fields is $u^\alpha=\{\psi_{ab}, \Pi_{ab}, \Phi_{iab},
H_a, \theta_a\}$, where Greek letters are used for indices that
enumerate the dynamical fields.

The hyperbolicity of a first-order evolution system, such as
Eq.~(\ref{e:FOSystem}), is determined by the properties of the
characteristic matrix $A^{k\,\alpha}{}_\beta$.  We define the left
eigenvectors $e^{\hat \alpha}{}_\alpha$ and their associated eigenvalues
$v_{(\hat\alpha)}$ of the characteristic matrix in the following way,
\begin{eqnarray}
e^{\hat\alpha}{}_{\beta}\,n_k A^{k\,\beta}{}_\alpha = v_{(\hat\alpha)}
e^{\hat\alpha}{}_\alpha,
\label{e:eigenproblem}
\end{eqnarray}
where $n_k$ denotes a spacelike unit vector; accented Greek
letters, $\hat\alpha$, ..., are used to enumerate distinct linearly
independent eigenvectors.  The eigenvalues, $v_{(\hat\alpha)}$, are
often referred to as the characteristic speeds of the system.  A
first-order evolution system is strongly hyperbolic at a point in
spacetime if there exists a complete set of eigenvectors for each
$n_k$ at that point.  In this case the matrix of eigenvector
components $e^{\hat\alpha}{}_\alpha$ is non-degenerate, i.e., $\det
e^{\hat\alpha}{}_\alpha\neq 0$.  The projections of the dynamical
fields onto the eigenvectors, $u^{\hat\alpha}=e^{\hat\alpha}{}_\alpha
u^\alpha$, provide an alternate complete set of dynamical fields,
which play an important role in strongly hyperbolic systems.  For
example, the characteristic fields, $u^{\hat\alpha}$, are those on
which appropriate boundary conditions must be placed for these
systems.

It is fairly straightforward to work out the characteristic
eigenvalues and eigenvectors, and the associated characteristic
fields, for the combined GH Einstein gauge-driver system:
\begin{eqnarray}
u^{\hat 0}_{ab}      &=& \psi_{ab},\label{e:Def_u0}\\
u^{{\hat 1}\pm}_{ab} &=& \Pi_{ab} \pm n^i \Phi_{iab}
                           -\gamma_2 \psi_{ab}
                           \pm n_a H_b\pm n_b H_a,
\label{e:Def_u1pm} \\
u^{\hat 2}_{iab}     &=& P_i{}^k \Phi_{kab},
\label{e:Def_u2}\\
u^{\hat 3}_{a}       &=& H_a,
\label{e:Def_u3}\\
u^{\hat 4}_{a}       &=& \theta_a + \eta H_a,\label{e:Def_u4}
\end{eqnarray}
where $P_i{}^k=\delta_i{}^k - n_i n^k$.  We see that the coupling
between the GH Einstein and gauge-driver systems increases the number
of characteristic fields, and also transforms the characteristic
fields of the pure GH Einstein system.  This means that the theory of
the boundary conditions for the GH Einstein system will have to be
completely re-examined. We also note that the co-vector $n_a$ is
a spatial unit normal, which is orthogonal to the timelike unit normal
$t_a$.  This implies that the spatial components of $n_a$ are the
usual components of the spatial normal co-vector $n_i$ while the time
component $n_t$ must be given by: $n_t = n_k N^k$.  These conditions
ensure that $t^a n_a =0$ and $n^a n_a=n^k n_k = 1$.

The characteristic speeds, $v_{(\hat\alpha)}$, associated with the
combined GH Einstein gauge-driver system are as follows: the fields
$u^{\hat 0}_{ab}$ have coordinate characteristic speed
$-(1+\gamma_1)n_kN^k$, the fields $u^{{\hat 1}\pm}_{ab}$ have speed
$-n_kN^k\pm N$, the fields $u^{\hat 2}_{iab}$ and $u^{\hat 3}_a$ have
speed $-n_kN^k$, and the fields $u^{\hat 4}_a$ have speed zero.  On
boundary points each characteristic field (computed with the outward
directed unit normal to the boundary $n_k$) must be supplied with a
boundary condition if and only if its associated characteristic speed
is negative.  The appropriate boundary conditions for the combined GH
Einstein gauge-driver system are discussed in some detail in
Appendix~\ref{s:BoundaryConditions}.

The inverse transformation between dynamical and characteristic fields
for the combined GH Einstein gauge-driver system is
\begin{eqnarray}
\psi_{ab}&=& u^{\hat 0}_{ab},\\
\Pi_{ab}&=& \half(u^{\hat 1+}_{ab}+u^{\hat 1-}_{ab})
+\gamma_2 u^{\hat 0}_{ab},\\
\Phi_{iab} &=& \half n_i(u^{\hat 1+}_{ab}-u^{\hat 1-}_{ab})
+ u^{\hat 2}_{iab} \nonumber\\
&&- n_i(n_a u^{\hat 3}_b + n_b u^{\hat 3}_a),\\
H_a &=& u^{\hat 3}_a,\\
\theta_a &=& u^{\hat 4}_a - \eta \,u^{\hat 3}_a.
\end{eqnarray}
Since this transformation is invertible, the combined first-order GH
Einstein gauge-driver evolution system is strongly hyperbolic. 

A first-order evolution system, Eq.~(\ref{e:FOSystem}), is called
symmetric hyperbolic, if there exists a symmetric positive definite
matrix on the space of dynamical fields, $S_{\alpha\beta}$, that
symmetrizes the characteristic matrices: $S_{\alpha\gamma}
A^{k\,\gamma}{}_\beta\equiv A^k_{\alpha\beta}=A^k_{\beta\alpha}$.
Symmetric hyperbolic systems provide a natural ``energy,'' $E=\int
S_{\alpha\beta}u^\alpha u^\beta d^{\,3}x$, and are better behaved than
strongly hyperbolic systems for initial-boundary value problems.
Symmetric hyperbolicity is therefore a desirable property for
gauge-driver systems to have.  It is fairly straightforward to show
that the combined GH Einstein gauge-driver system of
Eqs.~(\ref{e:Einstein}), (\ref{e:GaugeThetaEq}) and (\ref{e:GaugeHEq})
has a symmetrizer given by:
\begin{eqnarray}
&&\!\!\!\!\!
dS^2= S_{\alpha\beta}du^\alpha du^\beta,\nonumber\\
&& \quad\, = m^{ab}\left[
\Lambda^2_\psi m^{cd}d\psi_{ac}d\psi_{bd}
+\Lambda^2_H dH_a dH_b \right.\nonumber\\ 
&&\qquad\qquad\quad
\left.+ \Lambda^2_\theta(d\theta_a +\eta dH_a)(d\theta_b+\eta dH_b)\right]
\nonumber\\
&&+m^{ab}m^{cd}\left[g^{ij}\left(d\Phi_{iac}+2 g_{ia}dH_c\right)
\left(d\Phi_{jbd}+2g_{jb}dH_d\right)
\right.
\nonumber\\
&&\qquad\qquad\,\,
\left.+ \left(d\Pi_{ac}-\gamma_2d\psi_{ac}\right)
\left( d\Pi_{bd}-\gamma_2d\psi_{bd}\right) \right].
\label{e:GaugeDriverSymmetrizer}
\end{eqnarray}
This symmetrizer is positive definite as long as $m^{ab}$ is a
positive definite symmetric tensor, and the
(real) scalars $\Lambda_\psi$,
$\Lambda_H$, and $\Lambda_\theta$ are non-vanishing.  Therefore the
gauge-driver system of Eqs.~(\ref{e:Einstein}), (\ref{e:GaugeThetaEq})
and (\ref{e:GaugeHEq}) is symmetric hyperbolic.

%%%%%%%%%%%%%%%%%%%%%%%%%%%%%%%%%%%%%%%%%%%%%%%%%%%%%%%%%%%%%%%%%%%%%%%%%%%%%%
\section{Dual Coordinate Frames}
\label{s:DualCoordinateFrameEvolution}
%%%%%%%%%%%%%%%%%%%%%%%%%%%%%%%%%%%%%%%%%%%%%%%%%%%%%%%%%%%%%%%%%%%%%%%%%%%%%%

We have found that using two different coordinate systems
simultaneously is a very useful numerical technique, when performing
numerical evolutions of binary black-hole
spacetimes~\cite{Scheel2006}.  This method allows us to choose one set
of coordinates, $x^a$ thought of as ``co-moving,'' to track
(approximately) the motion of the black holes, and a second set,
$x^{\bar a}$ thought of as ``inertial,'' fixed (approximately) to a
non-rotating frame at infinity.  We evaluate the components of the
various dynamical fields using tensor bases defined by the inertial
$x^{\bar a}$ coordinates, while the evolution equations are solved for
those inertial-frame field components $u^{\bar\alpha}$ as functions of
the moving $x^a$ coordinates.  This use of dual coordinate frames
minimizes the size of the various field components and their time
derivatives better than any single-frame coordinate choice.

The single-frame GH Einstein gauge-driver equations, introduced in
Sec.~\ref{s:FirstOrderGaugeDriver}, written in terms of inertial-frame
quantities are given by
\begin{eqnarray}
&&\partial_{\bar t} H_{\bar a} - {\bar N}^{\bar k}\partial_{\bar k} H_{\bar a} 
= -\mu(H_{\bar a} - F_{\bar a}) + \theta_{\bar a},
\label{e:InertialGaugeH}
\\
&&\partial_{\bar t}\theta_{\bar a} + \eta\,{\bar N}^{\bar k}\partial_{\bar k} 
H_{\bar a} = -\eta\,\theta_{\bar a}.
\label{e:InertialGaugeTheta}
\end{eqnarray}
These equations, together with the inertial-frame representations of
the Einstein system, can be converted to dual-frame form in a
straightforward way using the prescription developed in
Ref.~\cite{Scheel2006}.  Under this recipe, a first-order evolution
system for inertial frame components, $u^{\bar \alpha}$,
\begin{eqnarray}
\partial_{\bar t} u^{\bar \alpha} + A^{\bar k\,\bar \alpha}{}_{\bar \beta}
\partial_{\bar k} u^{\bar \beta}= B^{\bar \alpha},
\label{e:FOSHSystem}
\end{eqnarray}
is converted into the dual-frame system
\begin{eqnarray}
\partial_{t} u^{\bar \alpha} + \left[\partial_{\bar t} x^i 
\delta^{\bar\alpha}{}_{\bar\beta}+\partial_{\bar k}x^i
A^{\bar k\,\bar \alpha}{}_{\bar \beta}\right]
\partial_{i} u^{\bar \beta}= B^{\bar \alpha},
\label{e:DualFrameFOSHSystem}
\end{eqnarray}
simply by changing independent variables: $\partial_{\bar t} =
\partial_t+\partial_{\bar t} x^{i}\partial_i$ and $\partial_{\bar k} =
\partial_{\bar k} x^{i}\partial_i$.  The quantities $\partial_{\bar
t}x^i\equiv \partial x^i/\partial {\bar t}$ and $\partial_{\bar
k}x^i\equiv \partial x^i/\partial x^{\bar k}$ are the non-trivial
parts of the Jacobian of the transformation relating the two
coordinate frames.  These coordinate transformations are assumed to be
given {\it a priori.}

The straightforward conversion of the GH Einstein gauge-driver system
from its inertial single-frame form, (\ref{e:InertialGaugeH}), and
(\ref{e:InertialGaugeTheta}), to dual-frame form may not always be the
most effective choice however.  The single-frame evolution equation
for $H_{\bar a}$, Eq.~(\ref{e:InertialGaugeH}), is designed to drive
$H_{\bar a}\rightarrow F_{\bar a}$ at fixed values of the inertial
coordinates.  A binary black-hole spacetime, however, can have rapid
time variations in the field components when evaluated at fixed
inertial coordinates, e.g., at points lying near the black-hole
trajectories.  The gauge-driver system will not be very efficient in
accurately enforcing the desired gauge under these very dynamical
conditions.  In contrast the moving coordinates, $x^a$, are chosen to
track (approximately) the motion of the holes, so the fields expressed
as functions of the moving coordinates are far less time dependent.  A
moving-frame version of the gauge-driver would therefore be more
effective enforcing the desired gauge, $H_{\bar a}=F_{\bar a}$, in
many situations.  In this case it makes sense to modify the evolution
equation for $H_{\bar a}$ in a way that ensures the moving-frame
components of $H_a$ are driven to the intended targets:
$H_{a}\rightarrow F_{a}$.  The appropriate moving-frame gauge driver
equations are simply Eqs.~(\ref{e:GaugeHEq}) and
(\ref{e:GaugeThetaEq}) interpreted now as moving-frame equations:
\begin{eqnarray}
&&\partial_t H_a - N^k\partial_k H_a = -\mu(H_a - F_a) + \theta_a,
\label{e:MovingGaugeH}
\\
&&\partial_t\theta_a + \eta\,N^k\partial_k H_a = -\eta\,\theta_a.
\label{e:MovingGaugeTheta}
\end{eqnarray}
It is straightforward to re-express these equations in terms of
inertial frame quantities:
\begin{eqnarray}
&&
\partial_{\bar t} H_{\bar a} 
- \bar N^{\bar k}\partial_{\bar k} H_{\bar a} = 
-\mu(H_{\bar a} - F_{\bar a}) + \theta_{\bar a}
\nonumber\\
&&\qquad
+\left(\partial_{\bar t} \partial_{\bar a} x^{a}-\bar N^{\bar k}
\partial_{\bar k}
\partial_{\bar a}x^{a}\right)\partial_{a}x^{\bar b}H_{\bar b},
\label{e:DualGaugeHEq}\\
&&\partial_{\bar t}\theta_{\bar a} 
+\partial_{t}x^{\bar k}\partial_{\bar k}\theta_{\bar a}
+ \eta\,\left(\bar N^{\bar k}+\partial_t x^{\bar k}\right)
\partial_{\bar k} H_{\bar a} = 
-\eta\,\theta_{\bar a}\qquad\nonumber\\
&&\qquad+
(\partial_{\bar t} \partial_{\bar a} x^{a}
+
\partial_t x^{\bar k}\partial_{\bar k}\partial_{\bar a} x^{a})
\partial_{a}x^{\bar b}
\theta_{\bar b}
\nonumber\\
&&\qquad+\eta (\bar N^{\bar k}+\partial_t x^{\bar k})
(\partial_{\bar k}\partial_{\bar a}x^{a})
\partial_a x^{\bar b}H_{\bar b},
\label{e:DualGaugeThetaEq}
\end{eqnarray}
where $\partial_{\bar a} = \partial_{\bar a} x^a \partial_a$
transforms the derivatives,  $H_{\bar a} =
\partial_{\bar a} x^a H_a$ and $\theta_{\bar a} = \partial_{\bar a}
x^a \theta_a$ transform the field components, and the
inertial-frame shift $\bar N^{\bar k}$ is related to the moving-frame
shift $N^k$ by
\begin{equation}
N^k = \left( \bar N^{\bar k} + \partial_t x^{\bar k} \right)
      \partial_{\bar k} x^k.
\end{equation}

In some circumstances it may be advantageous to apply the
inertial-frame version of the gauge driver,
Eqs.~(\ref{e:InertialGaugeH}) and (\ref{e:InertialGaugeTheta}), in one
region of spacetime, while applying the moving-frame version,
Eq.~(\ref{e:DualGaugeHEq}) and (\ref{e:DualGaugeThetaEq}), in another.
For example, in a binary black-hole simulation it might be appropriate
to impose the moving-frame version of the gauge driver in the very
dynamical region of spacetime near the black holes, while imposing the
inertial-frame version in the more quiescent asymptotic region far
from the holes.  Therefore, to accommodate this possibility we
introduce the following hybrid gauge-driver system that simply interpolates
between the two:
\begin{eqnarray}
&&
\partial_{\bar t} H_{\bar a} 
- \bar N^{\bar k}\partial_{\bar k} H_{\bar a} = 
-\mu(H_{\bar a} - F_{\bar a}) + \theta_{\bar a}
\nonumber\\
&&\qquad
+w\left(\partial_{\bar t} \partial_{\bar a} x^{a}-\bar N^{\bar k}
\partial_{\bar k}
\partial_{\bar a}x^{a}\right)\partial_{a}x^{\bar b}H_{\bar b},
\label{e:HybridDualGaugeHEq}\\
&&\partial_{\bar t}\theta_{\bar a} 
+w\partial_{t}x^{\bar k}\partial_{\bar k}\theta_{\bar a}
+ \eta\,\left(\bar N^{\bar k}+w\partial_t x^{\bar k}\right)
\partial_{\bar k} H_{\bar a} = 
-\eta\,\theta_{\bar a}\nonumber\\
&&\qquad+
w(\partial_{\bar t} \partial_{\bar a} x^{a}
+
\partial_t x^{\bar k}\partial_{\bar k}\partial_{\bar a} x^{a})
\partial_{a}x^{\bar b}
\theta_{\bar b}
\nonumber\\
&&\qquad+w\eta(\bar N^{\bar k}+\partial_t x^{\bar k})
(\partial_{\bar k}\partial_{\bar a}x^{a})
\partial_a x^{\bar b}H_{\bar b}.
\label{e:HybridDualGaugeThetaEq}
\end{eqnarray}
In these equations the smooth weight function $w$ is specified {\it a
priori}, with $w=0$ in the spacetime region where an inertial-frame
gauge driver is needed, and $w=1$ in the regions where a moving-frame
gauge driver is required.  The dual-frame version of this hybrid
gauge-driver system is obtained by combining these equations with the
inertial-frame Einstein system,
Eqs.~(\ref{e:NewPsiDot})--(\ref{e:NewPhiDot}), and using the
dual-frame conversion technique summarized in
Eqs.~(\ref{e:FOSHSystem}) and (\ref{e:DualFrameFOSHSystem}).

Since the hybrid gauge driver Eqs.~(\ref{e:HybridDualGaugeHEq}) and
(\ref{e:HybridDualGaugeThetaEq}) do not have the same principal parts
as their single-frame counterparts, we must consider again the
hyperbolicity of the combined GH Einstein plus hybrid gauge-driver
system.  Fortunately, we find that this system is still strongly
hyperbolic, and the characteristic fields are just
Eqs.~(\ref{e:Def_u0})--(\ref{e:Def_u4}) expressed in terms of
inertial-frame field components.  The characteristic speeds associated
with these fields are modified somewhat however: The fields $u^{\hat
0}_{\bar a\bar b}$ have inertial-coordinate characteristic speed
$-(1+\gamma_1)n_{\bar k}\bar N^{\bar k}$, the fields $u^{{\hat
1}\pm}_{\bar a\bar b}$ have speeds $-n_{\bar k}\bar N^{\bar k}\pm \bar
N$, the fields $u^{\hat 2}_{\bar i\bar a\bar b}$ and $u^{\hat 3}_{\bar
a}$ have speed $-n_{\bar k}{\bar N}^{\bar k}$, and $u^{\hat 4}_{\bar a}$ has
the speed $wn_{\bar k}\partial_t x^{\bar k}$.  In these expressions
$\bar N$ and $\bar N^{\bar k}$ refer to the inertial frame lapse and
shift respectively.  The co-moving-frame characteristic speeds are
obtained from the inertial-frame speeds by adding $-n_{\bar
k}\partial_t x^{\bar k}$.  This hybrid gauge driver system is also
symmetric hyperbolic with the same symmetrizer,
Eq.~(\ref{e:GaugeDriverSymmetrizer}), interpreted as an expression
in terms of inertial-frame field components.

%%%%%%%%%%%%%%%%%%%%%%%%%%%%%%%%%%%%%%%%%%%%%%%%%%%%%%%%%%%%%%%%%%%%%%%%%%%%%%
\section{Constraints}
\label{s:Constraints}
%%%%%%%%%%%%%%%%%%%%%%%%%%%%%%%%%%%%%%%%%%%%%%%%%%%%%%%%%%%%%%%%%%%%%%%%%%%%%%
This appendix investigates the constraints of the new GH Einstein
gauge-driver system.  These constraints and (somewhat surprisingly)
their evolution equations turn out to be identical to those of the
pure GH Einstein system.  This means that the constraint-preserving
boundary conditions derived for the pure GH Einstein system are also
appropriate for the combined GH Einstein gauge-driver system, although
care must be taken to enforce them on the correct characteristic
fields of the combined system.  This section presents the groundwork
for the detailed discussion of boundary conditions in
Appendix~\ref{s:BoundaryConditions}.

The primary constraint of the GH Einstein system is the gauge constraint,
${\cal C}_a$, which can be written in terms of the first-order
dynamical fields:
\begin{eqnarray}
{\cal C}_a &=& H_a + g^{ij}\Phi_{ija}
		 +t^b \Pi_{ba}
                  -\half g_a^i \psi^{bc}\Phi_{ibc}
		 -\half t_a \psi^{bc}\Pi_{bc}.\nonumber\\
\label{eq:OneIndexConstraint}
\end{eqnarray}
There are no extra constraints from the addition of the first-order
gauge-driver fields $H_a$ and $\theta_a$.  In the pure GH Einstein
system the gauge-source function $H_a$ is assumed to be a prescribed
function of the spacetime coordinates $x^a$ and the 4-metric
$\psi_{ab}$: $H_a=H_a(x,\psi)$.  In contrast $H_a$ is elevated to the
status of an independent dynamical field that is evolved according to
Eq.~(\ref{e:GaugeHEq}) in the combined GH Einstein gauge-driver system.
We need to determine, whether the evolution of the GH constraint
fields is affected by the introduction of this gauge-driver equation.
In addition we need to find the characteristic constraint fields to
determine what constraint preserving boundary conditions are needed
for the new combined system.

The basic GH Einstein system,
Eqs.~(\ref{e:NewPsiDot})--(\ref{e:NewPhiDot}) is, as before, just a
representation of the 4-dimensional co-variant Einstein equation:
\begin{eqnarray}
R_{ab}=\nabla_{(a}{\cal C}_{b)} 
- \gamma_0\bigl[t_{(a}{\cal C}_{b)} - \half \psi_{ab}t^c{\cal C}_c\bigr],
\end{eqnarray}
where $R_{ab}$ is the Ricci curvature, and $\nabla_a$ is the co-variant
derivative associated with $\psi_{ab}$.  Consequently the evolution
equation for ${\cal C}_a$ is determined by the Bianchi identities for
the 4-dimensional Ricci tensor, which can be
written as the second-order wave equation:
\begin{eqnarray}
0&=&\nabla^c\nabla_c {\cal C}_a 
-2\gamma_0 \nabla^b [\,t_{(b}{\cal C}_{a)}]
+ {\cal C}^{\,b}\nabla_{(a}{\cal C}_{b)}
-\half\gamma_0 \,t_a {\cal C}^{\,b}{\cal C}_{b}.
\nonumber\\
&&\label{e:CDamping}
\end{eqnarray}
This equation is identical to that obtained for the pure GH Einstein
system \cite{Lindblom2006}, because its derivation does not depend on
how the $H_a$ field is evolved.

Constraint preserving boundary conditions are designed to prohibit the
influx of constraint violations through the boundaries of the
computational domain.  In order to fix the incoming constraint fields,
the characteristic fields of the constraint evolution system must be
identified.  This is done most easily by transforming the second-order
constraint evolution Eq.~(\ref{e:CDamping}) to first-order form.  To
do this we introduce new constraint fields representing the first
derivatives of ${\cal C}_a$.  Thus we define new constraint fields
${\cal F}_a$ and ${\cal C}_{ia}$ that satisfy
\begin{eqnarray}
{\cal F}_a    &\approx& t^c\partial_c {\cal C}_a
=N^{-1}(\partial_t {\cal C}_a
                            -N^i \partial_i {\cal C}_a),\\
{\cal C}_{ia} &\approx& \partial_i {\cal C}_a,
\end{eqnarray}
where $\approx$ indicates equality up to terms proportional to the
gauge constraint ${\cal C}_a$ and the first-order GH Einstein
constraint ${\cal C}_{iab}\equiv \partial_i \psi_{ab}-\Phi_{iab}$.
The following expressions for ${\cal F}_a$ and ${\cal C}_{ia}$
accomplish this in a way that keeps the form of the constraint
evolution system as simple as possible:
\begin{eqnarray}
\label{eq:TimeDerivOfOneIndexConstraint}
{\cal F}_a &\equiv& 
\half g_a^i \psi^{bc}\partial_i \Pi_{bc}
- g^{ij} \partial_i \Pi_{ja}
- g^{ij} t^b \partial_i \Phi_{jba}
\nonumber \\ &&
+ g_a^i \Phi_{ijb} g^{jk}\Phi_{kcd} \psi^{bd} t^c
- \half g_a^i \Phi_{ijb} g^{jk}
  \Phi_{kcd} \psi^{cd} t^b
\nonumber \\ &&
+ \half t_a \psi^{bc} g^{ij} \partial_i \Phi_{jbc}
- \fourth  t_a g^{ij}\Phi_{icd}\Phi_{jbe}
   \psi^{cb}\psi^{de}
\nonumber \\ &&
- \half t_a g^{ij} g^{mn} \Phi_{imc} \Phi_{njd}\psi^{cd}
+ g^{ij} \Phi_{icd} \Phi_{jba} \psi^{bc} t^d
\nonumber \\ &&
+ \fourth  t_a \Pi_{cd} \Pi_{be} 
   \psi^{cb}\psi^{de}
- g^{ij} H_i \Pi_{ja}  
- t^b g^{ij} \Pi_{b i} \Pi_{ja}
\nonumber \\ &&
- \fourth  g_a^i \Phi_{icd} t^c t^d \Pi_{be}
  \psi^{be}
+ \half t_a \Pi_{cd} \Pi_{be}\psi^{ce}
  t^d t^b
\nonumber \\ &&
+ g_a^i \Phi_{icd} \Pi_{be} t^c t^b \psi^{de}
- \half g^{ij}\Phi_{icd} t^c t^d \Pi_{ja}
\nonumber \\ &&
- g^{ij}\Phi_{iba} t^b \Pi_{je} t^e
+ g_{a}^i \Phi_{icd} H_b \psi^{bc} t^d
+ t_a g^{ij} \partial_i H_j 
\nonumber \\ &&
+\gamma_2\bigl(g^{id}{\cal C}_{ida}
-\half  g_a^i\psi^{cd}{\cal C}_{icd}\bigr)
+\half  t_a g^{ij} H_i \Phi_{jcd}\psi^{cd}
\nonumber \\ &&
+ \half t_a \Pi_{cd}\psi^{cd} H_b t^b
- g^{ij} H_i \Phi_{jba} t^b
- g_a^i t^b \partial_i H_b
\nonumber \\ &&
- t_a g^{ij} \Phi_{ijc} H_d \psi^{cd}
,
\end{eqnarray}
\begin{eqnarray}
\label{eq:TwoIndexConstraint}
{\cal C}_{ia} &\equiv& g^{jk}\partial_j \Phi_{ika} 
- \half g_a^j\psi^{cd}\partial_j \Phi_{icd} 
- \half t_a \psi^{cd}\partial_i\Pi_{cd}
\nonumber\\&&
+ t^b \partial_i \Pi_{ba}
+ \partial_i H_a 
+ \half g_a^j \Phi_{jcd} \Phi_{ief} 
\psi^{ce}\psi^{df}
\nonumber\\&&
+ \half g^{jk} \Phi_{jcd} \Phi_{ike} 
\psi^{cd}t^e t_a
- g^{jk}g^{mn}\Phi_{jma}\Phi_{ikn}
\nonumber\\&&
+ \half \Phi_{icd} \Pi_{be} t_a 
                            \left(\psi^{cb}\psi^{de}
                      +\half\psi^{be} t^c t^d\right)
\nonumber\\&&
- \Phi_{icd} \Pi_{ba} t^c \left(\psi^{bd}
                            +\half t^b t^d\right)
\nonumber\\&&
+ \half \gamma_2 \left(t_a \psi^{cd}
- 2 \delta^c_a t^d\right) {\cal C}_{icd}.
\end{eqnarray}
We note that while ${\cal F}_a$ is defined as the time derivative of
${\cal C}_a$, the expression in
Eq.~(\ref{eq:TimeDerivOfOneIndexConstraint}) contains no time
derivatives.  The constraint fields are functions of the fundamental
dynamical fields of the system $u^\alpha$.  Any time derivatives of
the constraint fields are determined by the time derivatives of these
fundamental fields through the evolution equations of the system.
When the time derivatives of the expression for ${\cal C}_a$ in
Eq.~(\ref{eq:OneIndexConstraint}) are evaluated, and the time
derivatives of $\{\psi_{ab},\Pi_{ab},\Phi_{iab}\}$ are replaced with the
expressions from the basic GH Einstein system,
Eqs.~(\ref{e:NewPsiDot})--(\ref{e:NewPhiDot}), we find that the
occurrences of $\partial_t H_a$ cancel one another.  Thus the
expression for ${\cal F}_a$ does not depend on how $H_a$ is evolved,
and it is valid for both the pure GH Einstein system and the new
first-order gauge driver system.  To complete the GH constraint
evolution system we need to add the GH Einstein constraint ${\cal
C}_{iab}$,
\begin{eqnarray}
{\cal C}_{iab}=\partial_i\psi_{ab}-\Phi_{iab},
\label{e:C3def}
\end{eqnarray}
and the closely related ${\cal C}_{ijab}$, defined by
\begin{eqnarray}
{\cal C}_{ijab} &=& 2\partial_{[i}\Phi_{j]ab}
= 2\partial_{[j}{\cal C}_{i]ab}.
\label{eq:FourIndexConstraint}
\end{eqnarray}

The complete collection of constraints for the GH Einstein
gauge-driver evolution system is the set $c^I\equiv\{{\cal C}_{a},
{\cal F}_{a}, {\cal C}_{ia}, {\cal C}_{iab}, {\cal C}_{ijab}\}$
defined in Eqs.~(\ref{eq:OneIndexConstraint}),
(\ref{eq:TimeDerivOfOneIndexConstraint}),
(\ref{eq:TwoIndexConstraint}), (\ref{e:C3def}), and
(\ref{eq:FourIndexConstraint}).  (We use upper case Latin indices to
enumerate the constraint fields.)  The constraints $c^I$ depend on the
dynamical fields $u^\alpha=\{\psi_{ab}, \Pi_{ab},
\Phi_{iab},H_a,\theta_a\}$ and their spatial derivatives $\partial_k
u^\alpha$.  We have evaluated these constraint evolution equations
using the new GH Einstein gauge-driver system and have verified that
they can be written in the abstract form
\begin{eqnarray}
\partial_t c^I + A^{k\,I}{}_J(u)\partial_k c^J = B^I{}_J(u,\partial u)\, c^J,
\label{e:ConstraintEvolutionSystem}
\end{eqnarray}
where $A^{k\,I}{}_J$ and $B^I{}_J$ may depend on the dynamical fields
$u^\alpha$ and their spatial derivatives $\partial_k u^\alpha$.  The
evolution of the constraint fields $c^I$ turns out to be completely
determined by the GH Einstein
Eqs.~(\ref{e:NewPsiDot})--(\ref{e:NewPhiDot}) alone without any use of
the gauge-driver Eqs.~(\ref{e:GaugeThetaEq}) and
(\ref{e:GaugeHEq}). While the constraint fields ${\cal C}_a$, ${\cal
F}_a$ and ${\cal C}_{ia}$ depend on $H_a$ and $\partial_k H_a$, the
time derivatives of these constraints are determined without using the
evolution equation for $H_a$, Eq.~(\ref{e:GaugeHEq}).  There is a
remarkable cancellation between the explicit time derivatives of $H_a$
appearing in $\partial_t{\cal F}_a$ and $\partial_t {\cal C}_{ia}$,
and the time derivatives of $H_a$ introduced when the $\partial_t
\Pi_{ta}$ terms are replaced in these expressions using the GH
evolution Eq.~(\ref{e:NewPiDot}).  Thus the constraint evolution
system for the first-order gauge-driver system does not depend at all
on the gauge driver Eqs.~(\ref{e:GaugeThetaEq}) and
(\ref{e:GaugeHEq}).  This constraint evolution system is identical to
the pure GH Einstein constraint evolution system given in
Ref.~\cite{Lindblom2006}, and is both strongly and symmetric
hyperbolic.

Since the constraint evolution equations for the GH Einstein
gauge-driver system are identical to those of the pure GH Einstein
system, the characteristic constraint fields $c^{\hat I}$ are also
identical.    The boundary conditions needed to ensure no
influx of constraint violations will also be the same therefore.  As
we have seen in Appendix~\ref{s:Hyperbolicity} however, the characteristic
dynamical fields $u^{\hat\alpha}$ of the two systems are not the same,
so the detailed expressions for the needed boundary conditions in the
two systems will be different.  So we recall here the expressions for
the characteristic constraint fields $c^{\hat I}$ from
Ref.~\cite{Lindblom2006}:
\begin{eqnarray}
{c}^{{\hat 0}\pm}_a   &=&  {\cal F}_a  \mp n^k {\cal C}_{ka}, 
\label{e:chat0pm}\\ 
{c}^{\hat 1}_{a}     &=& {\cal C}_a, 
\label{e:chat1} \\
{c}^{\hat 2}_{ia}    &=& P^k{}_i {\cal C}_{ka}, 
\label{e:chat2}\\
{c}^{\hat 3}_{iab} &=& {\cal C}_{iab}, 
\label{e:chat3}\\
{c}^{\hat 4}_{ijab} &=& {\cal C}_{ijab}.
\label{e:chat4}
\end{eqnarray}
\noindent
The characteristic constraint fields $c^{{\hat 0}\pm}_a$ have
coordinate characteristic speeds $-n_lN^l\pm N$, the fields $c^{\hat
1}_a$ have speed $0$, the fields $c^{\hat 2}_{ia}$ and $c^{\hat
4}_{ijab}$ have speed $-n_lN^l$, and the fields $c^{\hat 3}_{iab}$
have speed $-(1+\gamma_1)n_lN^l$.  Boundary conditions must be placed
on the incoming characteristic dynamical fields $u^{\hat\alpha}$ that
(among other things) fix the incoming characteristic constraint fields
$c^{\hat I}$ to zero.  These (and other) needed boundary conditions
are discussed next in Appendix.~\ref{s:BoundaryConditions}.

%%%%%%%%%%%%%%%%%%%%%%%%%%%%%%%%%%%%%%%%%%%%%%%%%%%%%%%%%%%%%%%%%%%%%%%%%%%%%%
\section{Boundary Conditions}
\label{s:BoundaryConditions}
%%%%%%%%%%%%%%%%%%%%%%%%%%%%%%%%%%%%%%%%%%%%%%%%%%%%%%%%%%%%%%%%%%%%%%%%%%%%%%

A boundary condition is required for each characteristic field
$u^{\hat \alpha}$ of the GH Einstein gauge-driver system, at each
boundary point where the characteristic speed $v_{(\hat \alpha)}$
associated with that field is negative.

The characteristic fields $u^{\hat 0}_{ab}$, Eq.~(\ref{e:Def_u0}),
have speed $-(1+\gamma_1)n_kN^k$ and may require boundary conditions
at some boundary points.  Since the constraints and the constraint
evolution equations of the GH Einstein gauge-driver system are
identical to those of the pure GH Einstein system, we can employ the
same approach to constructing constraint preserving boundary
conditions.  The constraint characteristic field $c^{\hat 3}_{iab}$,
Eq.~(\ref{e:chat3}), is related to the characteristic field $u^{\hat
0}_{ab}$ by the expression,
\begin{eqnarray}
n^i c^{\hat 3}_{iab}\approx d_\perp u^{\hat 0}_{ab},
\label{e:U0Perp}
\end{eqnarray}
where $d_\perp u^{\hat \alpha}$ denotes the
characteristic projection of the normal derivatives of
$u^{\hat\alpha}$, i.e., $d_\perp u^{\hat\alpha}\equiv
e^{\hat\alpha}{}_\beta n^k\partial_k u^\beta$, 
with $e^{\hat\alpha}{}_\beta$ defined in
Eq.~(\ref{e:eigenproblem}).
Here (and throughout
this appendix) $\approx$ implies that algebraic terms and terms
involving tangential derivatives of the fields
(e.g. $P_i{}^k\partial_k u^{\alpha}$) have not been displayed.  We
note that the constraint field $c^{\hat 3}_{iab}$ has the same
characteristic speed as $u^{\hat 0}_{ab}$.  Hence a constraint
preserving boundary condition for $c^{\hat 3}_{iab}$ is needed
whenever $u^{\hat 0}_{ab}$ needs a boundary condition.  The identity
relating $u^{\hat 0}_{ab}$ to $c^{\hat 3}_{iab}$,
Eq.~(\ref{e:U0Perp}), provides the way to formulate this boundary
condition by prescribing the value of $d_\perp u^{\hat 0}_{ab}$.

A convenient way has been found \cite{Lindblom2006} to impose
constraint preserving boundary conditions for fields like $u^{\hat
0}_{ab}$ that are related to an incoming constraint field through an
expression like Eq.~(\ref{e:U0Perp}).  The characteristic projection
of the time derivatives of these fields $u^{\alpha}$, $d_t
u^{\hat\alpha} \equiv e^{\hat\alpha}{}_\beta \partial_t u^\beta$, are
set in the following way at the boundary,
\begin{eqnarray}
d_t u^{\hat\alpha} = D_t u^{\hat\alpha} + v_{(\hat\alpha)}
\bigl(d_\perp u^{\hat\alpha}-d_\perp u^{\hat\alpha}\bigr|_{BC}\bigr).
\label{e:BCMethod}
\end{eqnarray}
In this expression the terms $D_tu^{\hat\alpha}$ represent the
projections of the right sides of the evolution system,
Eqs.~(\ref{e:NewPsiDot})--(\ref{e:NewPhiDot}), (\ref{e:GaugeThetaEq}) and
(\ref{e:GaugeHEq}),
so the
characteristic projections of the evolution equations at non-boundary
points would simply be $d_t u^{\hat\alpha} = D_tu^{\hat\alpha}$.  The
term $d_\perp u^{\hat\alpha}\bigr|_{BC}$ is the value to which
$d_\perp u^{\hat\alpha}$ is to be fixed on the boundary.  This form of
the boundary condition replaces all of the $d_\perp u^{\hat\alpha}$
that appears in $D_t u^{\hat\alpha}$ with $d_\perp
u^{\hat\alpha}\bigr|_{BC}$.  Applying this method to the $u^{\hat
0}_{ab}$ field, we arrive at the desired constraint preserving
boundary condition for this field,
\begin{eqnarray}
d_t u^{\hat 0}_{ab} &=& D_t u^{\hat 0}_{ab} - (1+\gamma_1)n_j N^j n^k
c^{\hat 3}_{kab}.
\end{eqnarray}
This boundary condition is the same in the new GH Einstein
gauge-driver system as in the pure GH Einstein
system~\cite{Lindblom2006}.

The characteristic field $u^{\hat 2}_{iab}$, Eq.~(\ref{e:Def_u2}), has
speed $-n_kN^k$, and so this field may require a boundary condition on
some boundary points.  The constraint characteristic field $c^{\hat
4}_{ijab}$, Eq.~(\ref{e:chat4}), has the same characteristic speed,
and hence it is natural to use the boundary condition on $u^{\hat
2}_{iab}$ to prevent the influx of this constraint.  Conveniently,
there is an identity relating $u^{\hat 2}_{iab}$ and $c^{\hat
4}_{ijab}$:
\begin{eqnarray}
n^i c^{\hat 4}_{ikab} \approx  d_\perp u^{\hat 2}_{kab}.\label{e:U2Perp}
\end{eqnarray}
This identity is identical in the GH Einstein gauge-driver and
the pure GH Einstein systems~\cite{Lindblom2006}.  So we follow the
strategy of Eq.~(\ref{e:BCMethod}), and use the following constraint
preserving boundary condition for $u^{\hat 2}_{iab}$:
\begin{eqnarray}
d_t u^{\hat 2}_{kab}&=& D_t u^{\hat 2}_{kab}
-n_lN^l n^iP^j{}_k c^{\hat 4}_{ijab}.
\end{eqnarray}

The characteristic field $u^{\hat 3}_a$, Eq.~(\ref{e:Def_u3}), also
has speed $-n_kN^k$, and so it may require a boundary condition on
some boundary points.  We have identified two possibilities for this
boundary condition.  First, the $H_a$ field is part of the basic gauge
constraint Eq.~(\ref{eq:OneIndexConstraint}).  So one possible
boundary condition for $u^{\hat 3}_a$ is simply to enforce this
constraint on the boundary:
\begin{eqnarray}
u^{\hat 3}_a&=& -g^{ij}\Phi_{ija}
		 -t^b \Pi_{ba}
                  +\half g_a^i \psi^{bc}\Phi_{ibc}
		 +\half t_a \psi^{bc}\Pi_{bc}.\nonumber\\
&&\label{e:HBCconstraint}
\end{eqnarray}
Another possibility is to use a boundary condition on $u^{\hat 3}_a$
that enforces the desired gauge condition $H_a=F_a$ on the boundary:
\begin{eqnarray}
u^{\hat 3}_a = F_a.\label{e:HBCF}
\end{eqnarray}
These boundary conditions could be imposed as Dirichlet conditions in
the forms given above using penalty methods.  Alternatively, we could
impose these conditions using Bjorhus methods as a driver condition on
the boundary value of the time derivative of the characteristic field,
\begin{eqnarray}
d_t u^{\hat \alpha} &=& -\mu_B
(u^{\hat \alpha}- u^{\hat \alpha}|_{BC}).
\label{e:BoundaryDriver}
\end{eqnarray}
The parameter $\mu_B$ sets the timescale on which the boundary value
of $u^{\hat\alpha}$ is driven to its target value.  The Bjorhus 
version of the boundary condition in Eq.~(\ref{e:HBCconstraint})
is therefore,
\begin{eqnarray}
d_t u^{\hat 3}_a&=&-\mu_B {\cal C}_a,
\end{eqnarray}
while the Bjorhus version of Eq.~(\ref{e:HBCF}) is
\begin{eqnarray}
d_t u^{\hat 3}_a&=&-\mu_B (H_a-F_a).
\label{e:HDriverBC}
\end{eqnarray}
In most of our numerical tests, we find that the
Eq.~(\ref{e:HDriverBC}) version of this boundary condition is more
effective.

The characteristic field $u^{\hat 4}_a$, Eq.~(\ref{e:Def_u4}), has
characteristic speed 0 in the single-frame evolution system, and hence
does not need a boundary condition in that case.  In the dual-frame
system the characteristic speed changes to $w n_{\bar k}\partial_t
x^{\bar k}$, so this field might need a boundary condition under some
conditions.  We have generally chosen weight functions $w$ and
dual-frame maps $\partial_t x^{\bar k}$ that avoid the need for a
boundary condition on this field.  But if that can not be done, it is
probably best to choose the boundary value of $\theta_a$ so that
$\partial_t \theta_a =0$ on the boundary.  This condition leads to the
following Dirichlet type boundary condition for $u^{\hat 4}_a$:
\begin{eqnarray}
u^{\hat 4}_a = \eta H_a - N^k\partial_k H_a.
\end{eqnarray}

Boundary conditions are rarely needed for the $u^{\hat 1+}_{ab}$
fields, i.e., only when the boundary of the computational domain moves
outward at superluminal speeds.  In contrast boundary conditions are
almost always needed for the $u^{\hat 1-}_{ab}$ fields.  These boundary
conditions split naturally into three types that have been called
gauge boundary conditions, constraint-preserving boundary conditions,
and physical boundary conditions \cite{Lindblom2006,Rinne2007}.  These three
different types of boundary conditions are imposed on the parts of
$u^{\hat 1-}_{ab}$ selected by the three mutually orthogonal
projection tensors: 
\begin{eqnarray}
P^{(G)}_{ab}{}^{cd}&=&-\bigl[k_ak_bl^{(c}+k_a\delta_b{}^{(c}
+k_b\delta_a{}^{(c}\bigr]l^{d)},\\
P^{(C)}_{ab}{}^{cd}&=&\half P_{ab}P^{cd} - 2l_{(a}P_{b)}{}^{(c}k^{d)}
                       +l_al_b k^c k^d,\\
P^{(P)}_{ab}{}^{cd}&=& P_a{}^{(c}P_b{}^{d)}-\half P_{ab}P^{cd}.
\end{eqnarray}
In these expressions $k^a$ and $l^a$ represent the ingoing and outgoing null
vectors, respectively, that are related to the timelike and outgoing
spacelike unit normal vectors, $t^a$ and $n^a$, by:
\begin{eqnarray}
k^a&=&\frac{1}{\sqrt{2}}\Bigl(t^a-n^a\Bigr),\\
l^a&=&\frac{1}{\sqrt{2}}\Bigl(t^a+n^a\Bigr).
\end{eqnarray}
Similarly $g_{ab}$ represents the spatial 3-metric and $P_{ab}$
the projection onto the 2-dimensional spatial boundary
surface:
\begin{eqnarray}
g_{ab}&=&\psi_{ab}+t_at_b,\\
P_{ab}&=&\psi_{ab}+t_at_b-n_an_b.
\end{eqnarray}
Finally, we note that the projection tensors $P^{(G)}_{ab}{}^{cd}$,
$P^{(C)}_{ab}{}^{cd}$, and $P^{(P)}_{ab}{}^{cd}$ are complete in the
sense that:
\begin{eqnarray}
\delta_a{}^{(c}\delta_b{}^{d)}
=P^{(G)}_{ab}{}^{cd}+P^{(C)}_{ab}{}^{cd}+P^{(P)}_{ab}{}^{cd}.
\end{eqnarray}
We now discuss the boundary conditions appropriate for the
three independent projections of the $u^{\hat 1-}_{ab}$ fields.

%%%%%%%%%%%%%%%%%%%%%%%%%%%%%%%%%%%%%%%%%%%%%%%%%%%%%%%%%%%%%%%%%%%%%%%%%%%%%%
\subsection{Gauge Boundary Conditions}
\label{s:GaugeBoundaryConditions}
%%%%%%%%%%%%%%%%%%%%%%%%%%%%%%%%%%%%%%%%%%%%%%%%%%%%%%%%%%%%%%%%%%%%%%%%%%%%%%

The term gauge boundary conditions is used to describe the boundary
conditions on the $P^{(G)}_{ab}{}^{cd}$ projection of $u^{\hat
1-}_{ab}$~\cite{Lindblom2006}.  From the structure of the
$P^{(G)}_{ab}{}^{cd}$ projection tensor, we see that these are in
effect boundary conditions on the $u^{\hat 1-}_{ab}l^b$ fields.
Writing out the definition of $u^{\hat 1-}_{ab}$, we see that
\begin{eqnarray}
u^{\hat 1-}_{ab}l^b&=&\Pi_{ab}l^b-n^i\Phi_{iab}l^b -\gamma_2 l_a -n_a H_bl^b
-\frac{1}{\sqrt{2}}H_a.
\label{e:GaugeComponentsU1minus}\nonumber\\
\end{eqnarray}
The $u^{\hat 1\pm}_{ab}$ characteristic fields determine the time and
the spatial derivatives of $\psi_{ab}$ normal to the boundary.  So
these gauge boundary conditions on $u^{\hat 1-}_{ab}l^b$ can be
thought of as fixing the $\Pi_{ab}l^b$ components of $\Pi_{ab}$.
Previously the gauge boundary condition on $u^{\hat 1-}_{ab}l^b$ has
been set by freezing the value of this projection of the
characteristic field, $P^{(G)}_{ab}{}^{cd}d_t u^{\hat
1-}_{cd}=0$~\cite{Lindblom2006}, or by imposing a Sommerfeld-like
boundary condition on this projection of $u^{\hat
1-}_{ab}$~\cite{Rinne2006}.

Here we present a new gauge boundary condition for $u^{\hat
1-}_{ab}l^b$ obtained by setting the target boundary value of
$\Pi_{ab}t^b$ to the value it would have if the gauge constraint were
satisfied exactly.  The components of $\Pi_{ab}t^b$ enter the gauge
constraint, ${\cal C}_a$, through the identity:
\begin{eqnarray}
\Pi_{ab}t^b &=& \bigl(\delta_a{}^b-t_at^b\bigr)\bigl({\cal C}_b-H_b
-g^{ij}\Phi_{ijb}+\half g_b{}^i\psi^{cd}\Phi_{icd}\nonumber\\
&&\qquad\qquad\qquad+\half t_b g^{ij}\Pi_{ij}\bigr).
\label{e:AltGaugeConstraint}
\end{eqnarray}
So using Eq.~(\ref{e:AltGaugeConstraint}) we set:
\begin{eqnarray}
\Pi_{ab}t^b\bigl|_{BC}&=& 
\bigl(\delta_a{}^b-t_at^b\bigr)\bigl(-H_b
-g^{ij}\Phi_{ijb}+\half g_b{}^i\psi^{cd}\Phi_{icd}\nonumber\\
&&\qquad\qquad\qquad+\half t_b g^{ij}\Pi_{ij}\bigr)\nonumber\\
&=&\Pi_{ab}t^b - \bigl(\delta_a{}^b-t_at^b\bigr){\cal C}_b.
\end{eqnarray}
Using this expression in the equation for $u^{\hat 1-}_{ab}l^b$
in Eq.~(\ref{e:GaugeComponentsU1minus}), we find the expression
for the target boundary value of $u^{\hat 1-}_{ab}l^b$ to be:
\begin{eqnarray}
u^{\hat 1-}_{ab}l^b\bigl|_{BC}&=&u^{\hat 1-}_{ab}l^b-\frac{1}{\sqrt{2}}
\bigl(\delta_a{}^b-t_at^b\bigr){\cal C}_b.
\end{eqnarray}
This boundary condition can either be imposed as a Dirichlet condition
by penalty methods, or as a boundary-driver condition by Bjorhus
methods using Eq.~(\ref{e:BoundaryDriver}).  The Bjorhus version of this
new gauge boundary condition is:
\begin{eqnarray}
P^{(G)}_{ab}{}^{cd}d_t u^{\hat 1-}_{cd}&=&-\mu_B P^{(G)}_{ab}{}^{cd}\bigl(
u^{\hat 1-}_{cd}-u^{\hat 1-}_{cd}\bigl|_{BC}\bigr),\nonumber\\
&=& \frac{\mu_B}{\sqrt{2}}\Bigl(k_ak_bl^c+k_a\delta_b{}^c
+k_b\delta_a{}^c\Bigr)\nonumber\\
&&\qquad\times\bigl(\delta_c{}^d-t_ct^d\bigr){\cal C}_d.
\end{eqnarray}

%%%%%%%%%%%%%%%%%%%%%%%%%%%%%%%%%%%%%%%%%%%%%%%%%%%%%%%%%%%%%%%%%%%%%%%%%%%%%%
\subsection{Constraint Preserving Boundary Conditions}
\label{s:ConstraintBoundaryConditions}
%%%%%%%%%%%%%%%%%%%%%%%%%%%%%%%%%%%%%%%%%%%%%%%%%%%%%%%%%%%%%%%%%%%%%%%%%%%%%%

The term constraint preserving boundary conditions is used to describe
the boundary conditions on the $P^{(C)}_{ab}{}^{cd}$ projection of
$u^{\hat 1-}_{ab}$.  These boundary conditions have been constructed
to enforce the incoming components of the constraint characteristic
fields $c^{\hat 0}_a=0$, defined in Eq.~(\ref{e:chat0pm}), at
the boundary.  For the pure GH Einstein system, it was shown that
\begin{eqnarray}
&&\!\!\!\!\!\!
c^{\hat 0-}_a\approx 
\sqrt{2}\bigl[ k^{(c}\psi^{d)}{}_a -\half k_a\psi^{cd}\bigr]\nonumber\\
&&\qquad\qquad\qquad\times
n^k\partial_k\bigl(\Pi_{cd}-n^i\Phi_{icd}-\gamma_0\psi_{cd}\bigr).
\label{e:c0du1Old}
\quad\end{eqnarray}
For the case of the pure GH Einstein system, this gives an expression
for $c^{\hat 0-}_a$ in terms of the normal derivative of $u^{\hat
1-}_{ab}$, and so can be used to construct a boundary condition using
Eq.~(\ref{e:BCMethod}).  For the new GH Einstein gauge driver
considered here, the $u^{\hat 1-}_{ab}$ characteristic fields include
the additional terms $-n_{(a} H_{b)}$.  In the derivation of
Eq.~(\ref{e:c0du1Old}) from
Eqs.~(\ref{eq:TimeDerivOfOneIndexConstraint}) and
(\ref{eq:TwoIndexConstraint}), the terms involving spatial derivatives
of $H_a$ were treated as being prescribed, and so were counted as some
of the (many) algebraic terms not displayed.  Since $H_a$ has been
elevated to the status of a dynamical field in the new first-order
gauge-driver system, however, these terms can no longer be ignored.
It turns out that the $\partial_k H_a$ terms in
Eqs.~(\ref{eq:TimeDerivOfOneIndexConstraint}) and
(\ref{eq:TwoIndexConstraint}) give the following extra contributions
to Eq.~(\ref{e:c0du1Old}):
\begin{eqnarray}
&&\!\!\!\!\!\!
c^{\hat 0-}_a\approx 
\sqrt{2}\bigl[ k^{(c}\psi^{d)}{}_a -\half k_a\psi^{cd}\bigr]\nonumber\\
&&\qquad\times
n^k\partial_k\bigl(\Pi_{cd}-n^i\Phi_{icd}-\gamma_0\psi_{cd}
-n_c H_d - n_d H_c\bigr) \nonumber\\
&&\,\,\,\,\,
\approx  
\sqrt{2}\bigl[ k^{(c}\psi^{d)}{}_a -\half k_a\psi^{cd}\bigr]
d_\perp u^{\hat 1-}_{cd}.
\label{e:c0du1New}
\quad\end{eqnarray}
Using this expression in Eq.~(\ref{e:BCMethod}), we
then arrive at the needed boundary condition for the
constraint preserving components of $u^{\hat 1-}_{ab}$:
\begin{eqnarray}
&&
\!\!\!\!\!\!\!\!
P^{(C)}_{ab}{}^{cd}d_t u^{{\hat 1}-}_{cd} = P^{(C)}_{ab}{}^{cd}
      D_t u^{{\hat 1}-}_{cd} 
      +\sqrt{2}(N+n_jN^j)
\nonumber\\&&\qquad\qquad\quad
      \times\bigl[
	l_{(a} P_{b)}{}^{c} 
	-\half P_{ab}l^c 
	-\half l_a l_b k^c
	\bigr]c^{{\hat 0}-}_{c}.\qquad
\label{e:U1mBC}
\end{eqnarray}
These boundary conditions have the same form as those derived
previously for the pure GH Einstein system~\cite{Lindblom2006}.  Here
however, the characteristic field $u^{\hat 1-}_{ab}$ has a different
meaning, since it depends explicitly on the $H_a$ field in the
GH Einstein gauge-driver case.

%%%%%%%%%%%%%%%%%%%%%%%%%%%%%%%%%%%%%%%%%%%%%%%%%%%%%%%%%%%%%%%%%%%%%%%%%%%%%%
\subsection{Physical Boundary Conditions}
\label{s:PhysicalBoundaryConditions}
%%%%%%%%%%%%%%%%%%%%%%%%%%%%%%%%%%%%%%%%%%%%%%%%%%%%%%%%%%%%%%%%%%%%%%%%%%%%%%

The term physical boundary condition is used to describe the boundary
condition on the $P^{(P)}_{ab}{}^{cd}$ projection of $u^{\hat
1-}_{ab}$~\cite{Lindblom2006}.  This projection corresponds to the
transverse traceless components of the metric field, and so describes
the physical gravitational wave degrees of freedom of the system.  In
the vacuum region far away from compact sources, the gravitational-wave
degrees of freedom are described by the propagating components of the
Weyl curvature tensor.  The characteristic fields, $w^\pm_{ab}$,
representing these incoming and outgoing wave degrees-of-freedom
respectively of the Weyl tensor, are given by
\begin{eqnarray}
w^\pm_{ab}&=&P^{(P)}_{ab}{}^{cd}(t^e\mp n^e)(t^f\mp n^f)C_{cedf}.
\end{eqnarray}
It is straightforward to show that the incoming gravitational-wave
characteristic field $w^-_{ab}$ depends on the normal derivatives of
the dynamical fields at the boundary by the expression:
\begin{eqnarray}
w^-_{ab}&\approx& P^{(P)}_{ab}{}^{cd} n^k\partial_k(\Pi_{cd}-n^i\Phi_{icd}),
\end{eqnarray}
where $\approx$ denotes that algebraic terms and terms depending on
tangential derivatives of the dynamical fields are not shown.  The
derivation of this expression depends on the fact that the physical
projection $P^{(P)}_{ab}{}^{cd}$ annihilates terms like
$P^{(P)}_{ab}{}^{cd}\psi_{cd}=0$ and
$P^{(P)}_{ab}{}^{cd}n_{(c}H_{d)}=0$.  Therefore the principal part of
$w^{-}_{ab}$ depends on the normal derivative of $u^{\hat 1-}_{ab}$:
\begin{eqnarray}
w^-_{ab}&\approx& P^{(P)}_{ab}{}^{cd} d_\perp u^{\hat 1-}_{cd}.
\label{e:WeylCharacteristicField}
\end{eqnarray}
This is the same expression (up to terms proportional to constraints)
that is satisfied in the pure GH Einstein system
case~\cite{Lindblom2006}, where the gauge-source functions are
prescribed: $H_a=H_a(x,\psi)$.  But here the characteristic field
$u^{\hat 1-}_{ab}$ has a somewhat different meaning.  The lowest-order
physical boundary condition is designed to enforce the no-incoming
wave condition $w^-_{ab}=0$ at the boundary.  It does this by using
Eq.~(\ref{e:WeylCharacteristicField}) to replace the normal derivative
of $u^{\hat 1-}_{ab}$ which appears in the Einstein evolution equation
for $u^{\hat 1-}_{ab}$.  This boundary condition is enforced as a
Bjorhus condition on $u^{\hat 1-}_{ab}$,
\begin{eqnarray}
\!\!\!\!\!\!\!\!
P^{(P)}_{ab}{}^{cd}d_t u^{\hat 1-}_{cd} 
= P^{(P)}_{ab}{}^{cd}\bigl[D_t u^{\hat 1 -}_{cd}
-(N+n_kN^k)w^-_{cd}\bigr],
\end{eqnarray}
which is the same condition used in the pure GH Einstein system
case~\cite{Lindblom2006}.  Higher-order physical boundary conditions
have also been derived for the pure GH Einstein
system~\cite{Rinne2008b}, and these could be used, essentially without
modification, for the GH Einstein gauge-driver system as well.

%%%%%%%%%%%%%%%%%%%%%%%%%%%%%%%%%%%%%%%%%%%%%%%%%%%%%%%%%%%%%%%%%%%%%%%%%%%%%%%
% Acknowledgment
%%%%%%%%%%%%%%%%%%%%%%%%%%%%%%%%%%%%%%%%%%%%%%%%%%%%%%%%%%%%%%%%%%%%%%%%%%%%%%%

\acknowledgments We thank Mark Scheel and Keith Matthews for helpful
discussions concerning this work.  This work was supported in part by
grants from the Sherman Fairchild Foundation, by NSF grants
DMS-0553302, PHY-0601459, PHY-0652995, and by NASA grant NNX09AF97G.
Some of the computations for this investigation were performed at the
Jet Propulsion Laboratory using computers funded by the JPL Office of
the Chief Information Officer.

%%%%%%%%%%%%%%%%%%%%%%%%%%%%%%%%%%%%%%%%%%%%%%%%%%%%%%%%%%%%%%%%%%%%%%%%%%%%%%%
% Bibliography
%%%%%%%%%%%%%%%%%%%%%%%%%%%%%%%%%%%%%%%%%%%%%%%%%%%%%%%%%%%%%%%%%%%%%%%%%%%%%%%
\bibstyle{prd}
\bibliography{../References/References}

\end{document}